\documentclass[12pt]{iopart}

\usepackage{iopams}
\usepackage{cite}
\usepackage{graphicx}
 
\begin{document}
%January 8, 2020    Accepted for publication in Journal of Physics Communications 
% May 12,2019    (Submitted for publication) ArXiv 1905.06705

\title[Gravitational Three-Body Problem and a Modified Schr\"odinger Equation ]{Relation Between Solutions of the Schr\"odinger  Equation with Transitioning  Resonance Solutions of the Gravitational Three-Body Problem}

\author{Edward Belbruno}

%CQG Address
\address{ Department of Mathematics,  Yeshiva University,  New York, NY, USA; Department of Astrophysical Sciences, Princeton University, Princeton, NJ, USA  }

%CQG Mail to
\eads{\mailto{Edward.Belbruno@yu.edu}}

\begin{abstract}
It is shown that a class of  approximate resonance solutions in the three-body problem under the Newtonian gravitational force  are equivalent to quantized solutions of a modified Schr\"odinger equation for a wide range of masses that transition between energy states. In the macroscopic scale,
the resonance solutions are shown to transition from one resonance type to another through weak capture at one of the bodies, while in the  Schr\"odinger equation, one obtains quantized wave solutions  transitioning between different energies. The resonance transition dynamics provides a classical model of a particle moving between different energy states in the  Schr\"odinger  equation.  This methodology provides a connection between celestial and quantum mechanics. 
\end{abstract}
  
%\maketitle


\section{Introduction} 
\label{sec:intro}

 The purpose of this paper is to describe a mechanism to globally model the solutions of a modified  Schr\"odinger equation and how they transition between energy states,  with a special set of approximate resonance solutions to the classical gravitational Newtonian three-body problem, for a wide range of masses. 
 These resonance solutions transition from one resonance to another through the process of weak capture.

We consider a special version of the three-body problem that has proved to be useful in understanding  the complexities of three-body motion, in the macroscopic scale, going back to Poincar\'e \cite{SiegelMoser:1971}.  This is the circular restricted three body-problem, where the motion of one body, $P_0$, is studied as it moves under the influence of the gravitational field of $P_1, P_2$, assumed to move in mutual circular orbits of constant frequency $\omega$.
It is also assumed that the mass of $P_0$, labeled $m_0$, is negligible with respect to the masses of $P_1, P_2$, labeled $m_1, m_2$, respectively.  
In this paper, we will also assume that $m_2$ is much smaller than $m_1$, $m_2 \ll m_1$. For example, in the case of planetary objects, one can take $P_1, P_2$ to be the Earth, Moon, respectively, and $P_0$ to be a rock.  One can scale down $m_0, m_1, m_2$, as well as the relative distances between the particles, until the quantum scale is reached where the pure gravitational modeling is no longer sufficient.  When using a rotating coordinate system that rotates about  the center of mass of $P_1, P_2$ with constant frequency $\omega$, it is well known that Hamiltonian function for the motion of $P_0$ is time independent defining a conservative system (see Section \ref{sec:WeakCapture} ). 

For a general class of conservative systems, that includes, for example,  the restricted three-body problem considered here, it is known that such systems can be associated to the Schr\"odinger equation. As is described in Lanczos \cite{Lanczos:1949}, this can be done by computing the action function $S(x)$ for the motion of $P_0$, where $x$  is the position of $P_0$. \footnote{$S$ is obtained as the integral of the Lagrangian function $L(x,y), y = dx/dt$, over minimizing trajectories $x(t)$. } $S$ is a solution of the Hamilton-Jacobi partial differential equation associated to the restricted three-body problem. The $x(t)$ are orthogonal to the surface $S(x) =C$, for constant $C$. In this sense, the iso-$S$ surfaces locally determine the trajectories $x(t)$, that is for $t$ having sufficiently small variation. On the other hand, it can be shown that $S$ is the phase of wave solutions to the  Schr\"odinger equation. Thus, $S(x) = C$ is a wave surface. Conversely, starting with the $S$ satisfying the  Schr\"odinger equation, one obtains the Hamiltonian-Jacobi equation provided the Planck\rq{}s constant $\hslash \rightarrow 0$. This equivalence is local since in general $S$ can only be shown to locally exist as a solution to the Hamiltonian-Jacobi equation, which is the case for the three-body problem. For the full equivalence it is necessary that one restricts $\hslash \rightarrow 0$ that is not realistic. More significantly, this equivalence does not determine the global behavior of the solutions $x(t)$ and how they dynamically can model the transition of energy states for solutions of the  Schr\"odinger equation.  

This paper will use methods of dynamical systems to globally determine special resonance motions for $P_0$ that are shown to be equivalent to different energy states for a modified  Schr\"odinger equation and where the transition between energy states is equivalent to the process of weak capture in the three-body problem described in this paper. This result does not require the restriction of Planck\rq{}s constant, and doesn\rq{}t use Planck\rq{}s constant in the modeling.  The action function $S$  is not used. Due to the complexity of the motions described, it is seen that using the iso-$S$ surfaces to locally determine the global solutions would not seem feasible.

We describe a mechanism for the existence of a special family, $\mathfrak{F}$, of approximate resonance motions of $P_0$ about $P_1$, that transition from one resonance to another by the process of {\em weak capture} by $P_2$. This is a temporary capture defined in Section \ref{sec:Results}. These motions are approximately elliptical with frequency $\omega_1 \equiv \omega_1(m/n)$.  $\omega_1(m/n)(t)$   or equivalently $\omega_1(t)$, are  functions of the time $t$, where $\omega_1(m/n)$ is approximately equal to  the constant values $(m/n)\omega$, $m,n$ are positive integers.   That is, the period of motion of $P_0$ is approximately synchronized with the circular motion of $P_2$ about $P_1$, where in the time $P_0$ makes  approximately $n$ revolutions about $P_1$, $P_2$ makes $m$ revolutions about $P_1$.  The approximate resonance value of the frequency means that $|\omega_1(t) - (m/n)\omega| < \delta$, for a small tolerance $\delta$ as time varies for restricted time spans, described in Section \ref{sec:Results}.  When $P_0$ is moving 
on an approximate resonance orbit about $P_1$, it will eventually move away from this orbit and become captured temporarily about $P_2$, in weak capture. When $P_0$ escapes from this capture, it again moves about $P_1$ in another resonance elliptical orbit, with approximate resonance $m'/n'$. This process repeats either indefinitely, or ends when, for example, $P_0$ escapes the $P_1,P_2$-system. This also implies that the  approximate two-body energy $E_1$ of $P_0$  about $P_1$ can only take on a discrete set of  values, $E_1(m/n)(t)$  at each time $t$, which are approximately constant defined by the resonances.  This is stated as Result A in Section \ref{sec:Results} and as Theorem A in Section \ref{sec:WeakCapture}. The properties and dynamics of weak capture, and weak escape, are described Section \ref{sec:WeakCapture}. Comets can perform such resonance transitions(see Sections \ref{sec:Results} \ref{sec:WeakCapture}).

A modified Schr\"odinger equation is defined for the motion of $P_0$ about $P_1$, under the gravitational perturbation of $P_2$. This is first considered in the case of macroscopic masses.  It is given by, 
 \begin{equation}
-\frac{\sigma^2}{2\nu}\nabla^2 \Psi  + \bar{V} \Psi = E \Psi,
\label{eq:Schrodinger}
\end{equation} 
where $\nabla^2 \equiv \nabla \cdot \nabla$ is the Laplacian operator, $\bar{V}$ is an averaged three-body gravitational potential, $E$ is the energy, and $\nu$ is the reduced mass for $P_0, P_1$. $\sigma$ is a function that depends on $m_0, m_1$ and $G$, the gravitational constant. $\sigma$ replaces $\hslash = h/2\pi$, $h$ is Planck's constant that is in the classical Schr\"odinger equation.  In this case, for macroscopic values of the masses, since $P_0$ is not a wave, $\Psi$ is used to determine the probability distribution function, $|\Psi|^2$,  of locating $P_0$ near $P_1$ as a macroscopic body. We show in Section \ref{sec:SchroEqu} that $E$ can only take on the following approximate quantized values,
\begin{equation}
E _{\tilde{n}}= - { {4\sigma}\over{\tilde{n}^2} } ,
\label{eq:E-LeadingOrder}
\end{equation}
$\tilde{n} = 1,2,3, \ldots $.  As seen in Section \ref{sec:Results}, this implies that the frequency of $P_0$ takes a particularly simple form that is independent of any parameters. These frequencies have the approximate values, $8/\tilde{n}^3$.
$\Psi$ is explicitly computed in Section \ref{sec:SchroEqu}. $|\Psi|^2$ is shown to be exponentially decreasing as a function of the distance of $P_0$ from $P_1$.  The general solution, $\Psi$, of the modified Schr\"odinger equation is described in Result B in Section \ref{sec:Results}.
\medskip

A main result of this paper is that the quantized energy values $E_{\tilde{n}}$ correspond to a subset, $\mathfrak{U}$,  of the resonance orbit family, $\mathfrak{F}$, of $P_0$ about $P_1$. This is listed as  Result C in Section  \ref{sec:Results}. This provides a global equivalence of the solutions of the modified Schr\"odinger equation with the transitioning resonance solutions of the the three-body problem.
\medskip

As a final result, we show  is that the solution, $\Psi$, for the location of $P_0$ for the modified Schr\"odinger equation for the macroscopic values of the masses, can be extended into the quantum-scale.   This is summarized as Result D in Section \ref{sec:Results}. This gives a way to mathematically view the resonance motions in the quantum-scale, as an extension of the resonance solutions for macroscopic particles.  Other models, such as the classical Schr\"odinger and  Schr\"odinger-Newton equations are given in latter sections.

The results of this paper are described in detail and summarized in Section \ref{sec:Results}. This section contains the main findings of this paper. Additional details, derivations, and proofs are contained in Section \ref{sec:WeakCapture} for weak capture and in Section \ref{sec:SchroEqu} for the modified Schr\"odinger equation. 
\medskip

\section{Summary of Results, Definitions and Assumptions}
\label{sec:Results}
\medskip

In this section we elaborate on the results described in the Introduction. The first set of results pertain to
a family of resonance orbits about $P_1$ obtained from the three-body problem and the second set of results pertain to finding these orbits using a modified Schr\"odinger equation.
\medskip\medskip

\subsection{ Resonance Orbits in the Three-Body Problem and Weak Capture}
\medskip\medskip

The motion of $P_0$ is defined for the circular restricted three-body problem described in the Introduction. It is sufficient to use the planar version of this model, without loss of generality for the purposes of this paper, where $P_0$ moves in the same plane of motion as that of the uniform circular motion of $P_1, P_2$ of constant frequency $\omega$  (see Section \ref{sec:WeakCapture}).  The macroscopic masses satisfy, $m_2/m_1 \ll 1$ and the mass of $m_0$ is negligibly small so that $P_0$ does not gravitationally perturb $P_1, P_2$, but $P_1, P_2$ perturb the motion of $P_0$.  We consider an inertial coordinate system, $(X_1, X_2) \in \mathbb{R}^2$, whose origin is the center of mass of $P_1, P_2$.  

The differential equations for $P_0$ are given by the classical system
\begin{equation}
\ddot{X }= \Omega_{{X}}({X}, t),  
\label{eq:DE3Body}
\end{equation}
where $X = (X_1, X_2) \in \mathbb{R}^2$, $t \in \mathbb{R}^1$, ${^\cdot} \equiv {d\over{dt}}$, $\Omega_{X} \equiv (\Omega_{X_1}, \Omega_{X_2})$ ($\Omega_{X} \equiv \partial \Omega/\partial X$) and 
\begin{equation}
  \Omega =    {{G m_1}\over{r_1(t)}} + {{G m_2}\over{r_2(t)}} 
\label{eq:Omega}
\end{equation}
where $r_1(t) = |X - a_1(t)|$, $r_2 = |X - a_2(t)|$, $|\cdot|$ is the standard Euclidean norm.  The mutual circular orbits of $P_1, P_2$ are given by $a_1(t) =  \rho_1(\cos\omega_a t, \sin \omega_a t)$,  $a_2(t) =  -\rho_2(\cos\omega_b t, \sin \omega_b t)$, with constant circular frequencies  $\omega_a, \omega_b$ of $P_1, P_2$, respectively. We have divided both sides of (\ref{eq:DE3Body}) by $m_0$ and then took the limit as $m_0 \rightarrow 0$. It is well known that the solutions of the circular restricted three-body problem for $P_0$ accurately model the motion of $P_0$ in the general three-body problem for circular initial conditions for $P_1, P_2$, with $m_0$ kept positive and negligibly small.  
\medskip

\noindent
It is noted that all solutions $\xi(t) = (X(t), \dot{X}(t)) \in \mathbb{R}^4$ considered in this study will be $C^{\infty}$ in both $t$ and initial conditions, $\xi(t_0) = (X(t_0), \dot{X}(t_0))$ at an initial time $t_0$. We refer to $C^{\infty}$ as {\em smooth} dependence. More exactly, this means that all derivatives of $\xi(t)$ with respect to $t$ of all orders are continuous and all partial derivatives of $\xi(t,\xi(t_0))$ with respect to $X_1(t_0), X_2(t_0), \dot{X}_1(t_0), \dot{X}_2(t_0)$, of all orders, are continuous. 
\medskip

\noindent
Although $m_0$ is taken in the limit to be $0$ in the definition of the differential equations for the motion of $P_0$, we will assume it is non-zero but still negligible in mass with respect $P_1, P_2$,  $m_0 \gtrapprox 0$, in all equations that follow.     
\medskip
 
We transform to a $P_1$-centered coordinate system for the restricted three-body problem. In this system, $P_2$ moves about $P_1$ at a constant distance $\beta$, with constant circular frequency $\omega= \sqrt{G(m_1+m_2)/\beta}$. Before stating our first result, two definitions are needed. 
\medskip\medskip

  Assuming $m_2$ is much smaller than $m_1$, when $P_0$ moves about $P_1$ with elliptic initial conditions at an initial time $t=t_0$ , this elliptic motion will be slightly perturbed by $P_2$ \footnote{By the Kolmogorov-Arnold-Moser Theorem, the motion will stay approximately elliptic for all time for many initial conditions \cite{SiegelMoser:1971} } .  Let $a_1$  be the semi-major axis of $P_0$ with respect to $P_1$.  As a function of time, $a_1$ will vary. If $m_2=0$, then $a_1$ is constant since $P_0$ will move on a pure ellipse. If $m_2$ is  small, then  $P_0$ moves in a nearly elliptic orbit about $P_1$, and  $a_1(t)$ will be nearly constant for restricted time spans. This orbital element, along with the eccentricity, $e_1(t)$, true anamoly, $\theta_1(t)$,  and other orbital elements, can be calculated for each $t$ using the variational differential equations obtained from (\ref{eq:DE3Body}).(see \cite{Szebehely:1967},\cite{Pollard:1976}, \cite{Moulton:1970}. \cite{Stiefel:1971}). These are referred to as {\em  osculating elements}.  $e_1$ will likewise be nearly constant for a nearly elliptical orbit of $P_0$ anout $P_1$. 
\medskip

\noindent
The variation of   the frequency $\omega_1(t)$ can be obtained from $a_1(t)$:  The osculating two-body period, $T_1$, of $P_0$ is explicitly related to $a_1(t)$ by Kepler\rq{}s Third Law,  $a_1^3= (2\pi)^{-2}T_1^{2}G(m_0 + m_1) $, and $\omega(t) = T_1^{-1}$.
\medskip\medskip\medskip

\noindent
{\em Definition 1} \hspace{.05in}  An approximate resonance orbit, $\Phi_{m/n}(t)$, of $P_0$ moving about $P_1$ in a $P_1$-centered coordinate system, $Y = (Y_1, Y_2)$, as a function of $t$ in resonance with $P_2$, is an approximate elliptical orbit of frequency $\omega_1 = \omega_1(t)$, where  $\omega_1 \approx (m/n)\omega$. $m, n$ are positive integers. Thus, $\omega_1$ is approximately constant as time varies.  In phase space, $(Y, \dot{Y}) \in \mathbb{R}^4$, $\Phi_{m/n}(t)= (Y_1(t), Y_2(t), \dot{Y}_1(t), \dot{Y}_2(t))$.  $\Phi_{m/n}(t)$ has a period $T_1 = \omega_1^{-1}$, approximately constant. $T_1 \approx (n/m)T$, $T$ is the constant circular period of $P_2$ about $P_1$, $T = \omega^{-1}$. 
 For notational purposes, we refer to an approximate resonance orbit as a resonance orbit for short. A resonance orbit with $\omega_1 \approx (m/n){\bf \omega}$ is also referred to as a $n \mathbin{:} m$ resonance orbit.  ({\em Nearly resonant}  motion, related to approximate resonance motion,  is described in \cite{Naidon:2017}.)
\medskip\medskip

\noindent
The term 'approximate' in Definition 1 means to within a small tolerance, $\mathcal{O}(\delta)$,  $\delta = m_2/m_1 \ll 1$. $\mathcal{O}(\delta)$ is a function of time, $t$, and smooth in $t$.  An approximate elliptic orbit means that the variation of the orbital parameters($\omega_1$, $a_1$, $e_1$) of $\Phi(t) = (Y(t), \dot{Y}(t))$, with respect to $P_1$ in a $P_1$-centered coordinate system, will slightly vary by $\mathcal{O}(\delta)$ due to the gravitational perturbation of $P_2$. The two-body energy of $\Phi_{m/n}(t)$ with respect to $P_1$ is labeled, $E_1(m/n)$ {\bf = $E_1(m/n)(t)$}  , which is approximately constant.   Note that the general Kepler energy $E_1(t)$ along a trajectory $(Y(t), \dot{Y}(t))$ is given by (\ref{eq:KepEng1}).
\medskip

\noindent
Thus, $\omega_1 \approx (m/n)\omega$ is equivalent to $ \omega_1 = (m/n)\omega + \mathcal{O}(\delta)$. $a_1, e_1$ likewise vary within a variation $\mathcal{O}(\delta)$.  The variations $\mathcal{O}(\delta)$ are all different functions for the different parameters, but the same notation is used.   The tolerance on these orbital parameters is valid for finite times. We assume that $t$ varies over finite time spans. Thus, for a given variable, say $\omega_1(t)$, if $\epsilon_1>0$ is a given number, and $t \in [t_0, t_1]$, $t_1 > t_0$, $m_2$ can be taken small enough so that $|\mathcal{O}(\delta)| <  \epsilon_1$

\medskip

\noindent For a resonance orbit to be well defined, it is assumed that $m_2 > 0$. If $m_2 =0$,  then even though $\omega$ is defined, $P_2$ no longer exists. Thus, we assume $m_2 > 0$ throughout this paper, unless otherwise indicated. This assumption is also necessary for the definition of weak capture.
\medskip\medskip\medskip

We define 'weak capture' of $P_0$ about $P_2$.  In this case, we change to a $P_2$-centered coordinate system. This type of capture is discussed in Section \ref{sec:WeakCapture}.  Weak capture is where the two-body energy, $E_2$, of $P_0$ with respect to $P_2$ is temporarily negative.  It is used to define an interesting region about $P_2$ described in the next section called the weak stability boundary.  Chaotic motion occurs on and near this region.
\medskip\medskip

\noindent
{\em Definition 2} \hspace{.05in}  $P_0$ has {\em weak capture} about $P_2$, in a $P_2$-centered coordinate system, $Z =(Z_1, Z_2)$, at a time $t_0$ if the two-body energy, $E_2$, of $P_0$ with respect to $P_2$, is negative at $t_0$ and for a finite time after where it becomes positive($P_0$ escapes). More precisely, $E_2$ is given by
\begin{equation}                        
E_2 =  {1\over2} |{\dot{Z}}|^2 - {G(m_0 + m_2)\over r_2},  
\label{eq:KepEng}
\end{equation}
$r_2 = |{Z}| > 0$. 
Let $\mathcal{Z}(t) = (Z_1(t), Z_2(t), \dot{Z}_1(t), \dot{Z}_2(t)) $ be a solution for the differential equations (\ref{eq:DE3Body}) in $P_2$-centered coordinates for $t \ge t_0$. $P_0$ is weakly captured at $t_0$  if  $E_2(\mathcal{Z}(t)) < 0$ for $t_0 \leq t < t_1$, $t_0 < t_1$, $E_2(\mathcal{Z}(t_1)) = 0$, $E_2(\mathcal{Z}(t)) \gtrsim 0$ for $t \gtrsim t_1$.  After $P_0$ leaves weak capture at $t_1$, we say that $P_0$ has {\em weak escape} from $P_2$ at $t=t_1$.  Weak capture in backwards time from $t_0$ is similarly defined.
\medskip

\noindent
$P_0$ is {\em captured} at a point $(Z_1(t), Z_2(t))$ of a trajectory $\mathcal{Z}(t)$ at $t^*$ if  $E_2(\mathcal{Z}(t^*)) < 0$.  Capture at a point need not imply weak capture, in forward time, since $P_0$ could be captured for all time $t > t^*$.
\medskip\medskip

\noindent
This dynamics is summarized in Result A and proven in Section \ref{sec:WeakCapture}, where it is formulated more precisely as Theroem A. 
\medskip\medskip\bigskip

\noindent
{\bf Result A}
\medskip\medskip

\noindent
{\em Weak capture of $P_0$ about $P_2$ at a time $t_0$ yields resonance motion of $P_0$ about $P_1$, which repeats yielding a family, $\mathfrak{F}$, of resonance orbits.  More precisely,

\noindent
Assume $P_0$ is weakly captured by $P_2$ at time $t=t_0$, then 
\medskip

\noindent
(i)  As $t$ increases from $t_0$, $P_0$ first escapes from
$P_2$ and then $P_0$ moves onto a resonance orbit, $\Phi_{m/n}(t)$, about $P_1$. $P_0$ performs a finite number of cycles about $P_1$ until it eventually moves again to weak capture by $P_2$, where the process continues and $P_0$ moves onto another resonance orbit.  When $P_0$ moves from  $\Phi_{m/n}(t)$ to another resonance orbit, $\Phi_{m'/n'}(t)$, $m',n'$ may or may not equal $m,n$.  In general, a sequence of resonance orbits is obtained, $\{ \Phi_{m/n}(t), \Phi_{m'/n'}(t), \ldots \}$.  The process stops when $P_0$ escapes the $P_1P_2$-system, collides with $P_2$ or moves away from a resonance frequency.   This set of resonance orbits forms a family, $\mathfrak{F}$, of orbits, that depend on the initial weak capture condition.
\medskip

\noindent
(ii) When $P_0$ moves onto a sequence of resonance orbits about $P_1$ as described in (i), then a discrete set of energies are obtained, 
$\{ E_1(m,n), E_1(m',n'), \ldots\}$.}
\medskip\medskip\medskip
\medskip\medskip\medskip

\begin{figure}
\centering
	\includegraphics[width=0.70\textwidth, clip, keepaspectratio]{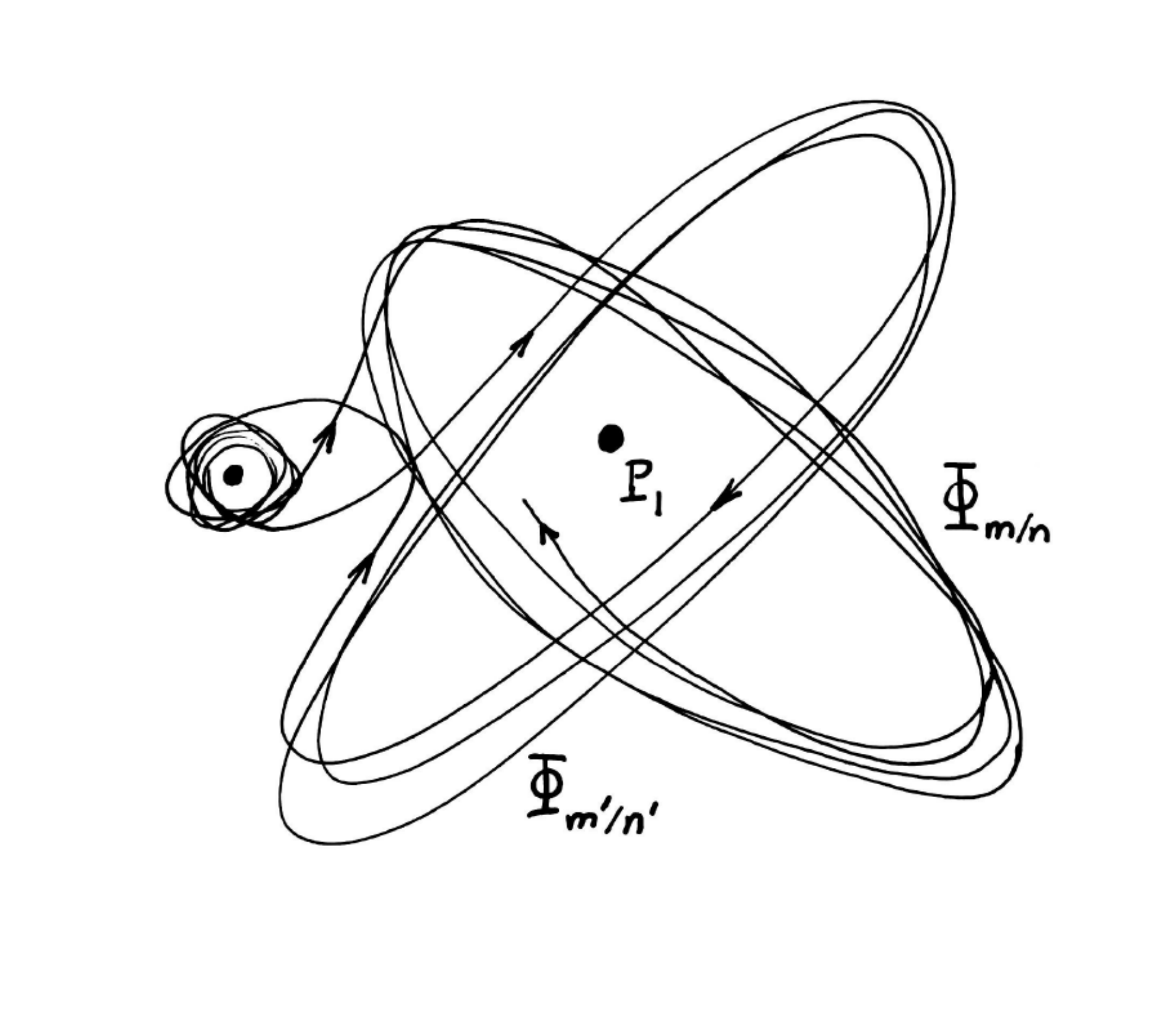}
%\centerline{\includegraphics{Fig1}}
\caption{Transitioning from one resonance orbit, $\Phi_{m/n}$, to another, $\Phi_{m'/n'}$, about $P_1$ through weak capture of $P_0$ near $P_2$.}
\label{fig:Fig1}
\end{figure}

\noindent
The transitioning of $\Phi_{m/n}(t)$ to $\Phi_{m'/n'}(t)$ is shown in a sketch in Figure \ref{fig:Fig1}.
\medskip\medskip

\noindent
The proof of Theorem A is given in detail in Section \ref{sec:WeakCapture}.   
\medskip\medskip

Applications of Theorem A a to comet motions and numerical simulations is described in Section \ref{sec:WeakCapture}.   
\medskip\medskip

A key result obtained in Section \ref{sec:WeakCapture} is,
\medskip\medskip

\noindent
{\bf Lemma A} \hspace{.05in} The frequency $ \omega_1(m/n){\bf(t)}$ of $\Phi_{m/n}(t)$ is given by,
\begin{equation}
\omega_1(m/n) =  (m/n)\omega + \mathcal{O}(\delta),
\label{eq:ResFrequenciesSec2}
\end{equation}
\medskip
where $\mathcal{O}(\delta)(t)$ is smooth in $t$. 
\medskip

(i) is proven in Section \ref{sec:WeakCapture} as Theorem A. We prove (ii):
\medskip\medskip\medskip

\noindent
 Consider the general two-body energy of $P_0$ with respect 
to $P_1$.  It is given by,
\begin{equation}
E_1 \equiv E_1(Y,  \dot{Y}) = {1\over2} |\dot{Y}|^2 - {{G (m_0+m_1)}\over r_1}  ,  
\label{eq:KepEng1}
\end{equation}                        
where $Y = (Y_1, Y_2)$, $r_1 = |Y|$, are inertial $P_1$-centered coordinates. $E_1$ can be written as $E_1 = -G(m_0+m_1)/(2a_1)$, \cite{Belbruno:2004}. Using Kepler's third law 
relating the period, $T_1$, to $a_1$, implies,   
$E_1 = -A\omega_1^{2/3}$, $A = (1/2)(2 \pi G(m_0 + m_1))^{2/3} $.  (\ref{eq:ResFrequenciesSec2}) implies $E_1$ can be written as,
\begin{equation}
E_1(\Phi_{m/n}(t)) \equiv E_1(m/n) = -[(m/n)\omega]^{2/3}A + \mathcal{O}(\delta),
\label{eq:ResE}
\end{equation}
where the remainder is smooth in $t$. (\ref{eq:ResE}) proves (ii).  It is noted that in this case, $E_1 \approx  -[(m/n)\omega]^{2/3}A$. 
\medskip\medskip\medskip

It is noted that $E_1$ can be written in an equivalent form to (\ref{eq:KepEng1}) by multiplying both sides of (\ref{eq:KepEng1}) by the reduced mass,  $\nu = m_0m_1(m_0+m_1)^{-1}$ , yielding 
\begin{equation}
\tilde{E}_1 \equiv \nu E_1 =   {\nu\over{2}} |\dot{Y}|^2 - {{G m_0m_1}\over r_1}  ,  
\label{eq:KepEng2}
\end{equation}                        
which implies, $\tilde{E}_1 = -Gm_0m_1/(2a_1).$ This scaled energy is more convenient to use when the modified Schr\"odinger equation is considered. To obtain the corresponding two-body differential equations with (\ref{eq:KepEng2}) as an integral, one multiplies the differential equations associated to (\ref{eq:KepEng1}) by $\nu$. The solutions are the same for both sets of differential equations.\footnote{(\ref{eq:KepEng1}) is an integral for $\ddot{Y} = -G(m_0+m_1)Yr_1^{-3}$, and (\ref{eq:KepEng2}) is an integral for $\nu\ddot{Y} = -Gm_0m_1Yr_1^{-3}$.}  Thus, it is seen that
$\tilde{E}_1 = \nu E_1 = -\nu A \omega_1^{2/3}$.  Setting $\sigma = \nu A$ implies,
\begin{equation}
 \tilde{E}_1 = -\sigma\omega_1^{2/3}, 
\label{eq:Planck2}
\end{equation}
\begin{equation}
\sigma = (1/2)(2 \pi G)^{2/3}m_0m_1(m_0 + m_1)^{-1/3}. 
\label{eq:Sigma}
\end{equation}
$\omega_1$  remains the same since the solutions haven't changed.  This defines the function $\sigma$ that plays a key role in this paper.
\medskip\medskip\medskip

\subsection{A Modified Schr\"odinger Equation: Macroscopic Scale }
\medskip\medskip

\noindent
       {\em Analogy A:} \hspace{.1in}    It is noted that Equation \ref{eq:Planck2} has a form similar to the Planck-Einstein relation for quantum mechanics for the energy, $\mathcal{E}$, of a photon, $ \mathcal{E} = h \lambda$, where $\lambda$ is the frequency of the photon. $\sigma$ is analogous to $h$ and $\tilde{\omega_1}^{2/3}$ is analogous to $\lambda$. There is another analogy for the case of an electron, $P_0$, moving about an atomic nucleus: When $P_0$ changes from one orbital to another, the energy of the photon absorbed or emitted is given by $\Delta E = h \lambda$, where $\Delta E = E_1 - E_2$, where $E_i$ is the energy of $P_0$ in the $i$th orbital, $i=1,2$. This process is analogous to a macroscopic particle $P_0$ changing from from one resonance orbit $\Phi_{m/n}(t)$ in $\mathfrak{F}$ to another, $\Phi_{m'/n'}(t)$, through weak capture and escape, where $\Delta \tilde{E}_1 = \tilde{E}_1(m/n)-\tilde{E}_1(m'/n')$. $\Delta \tilde{E}_1$ is analogous to $\Delta E$.
\medskip\medskip
\medskip
                                                                                     
\noindent
 We consider a  modified  Schr\"odinger equation (\ref{eq:Schrodinger}). This Schr\"odinger equation differs from the classical one by replacing $\hslash$ by the function $\sigma =\sigma(m_0, m_1, G)$ and $V$ by a three-body potential $\hat{V}$ derived from the circular restricted three-body problem. The choice of $\sigma$ is motivated by Analogy A.  In this modeling, the masses $m_0, m_1, m_2$ and distance between them is assumed to not be in the quantum scale. Under this assumption, $\Psi$ does not represent a wave motion and is used to measure the probability of locating $P_0$ at a distance $r_1$ from $P_1$.  
\medskip\medskip

The potential $\bar{V}$ in (\ref{eq:Schrodinger}), is derived from the restricted three-body problem modeling for the motion of $P_0$.  In inertial coordinates, $Y =(Y_1, Y_2)$ centered at $P_1$, $P_2$ moves about $P_1$ on a circular orbit, $\gamma(t)$ of constant radius $\beta$ and circular frequency $\omega = \sqrt{G(m_1+m_2)/\beta}$,  
$\gamma(t) = \beta(\cos \omega t, \sin \omega t)$.  The potential for $P_0$ is given by
\begin{equation}
\hat{V}  =  V_1 + V_2 ,
\label{eq:GeneralPotential}
\end{equation}
\begin{equation}
V_1 = -\frac{Gm_0m_1}{r},  \hspace{.2in}   V_2 = - \frac{Gm_0m_2}{r_2}, 
\label{eq:TwoBodyPotentials}
\end{equation}
where $r = |Y|, r_2 = |Y - \gamma(t)|$. For simplicity of notation, we have replaced the symbol $r_1$ by $r$ ($r_1 \equiv r)$.  

We replace $V_2$ by an approximation given by an averaged value of 
$V_2$ over one cycle of $\gamma(t)$, $t \in [0, 2 \pi/\omega]$,
\begin{equation}
\bar{V}_2 = -\frac{Gm_0m_2\omega}{2\pi} \int_0^{\frac{2\pi}{\omega}} \frac{dt}{\sqrt{r^2 + \beta^2 -2\beta(Y_1 \cos \omega t + Y_2 \sin \omega t)}}.
\label{eq:AveragedPotentialSec2}
\end{equation}

It is proven in Section \ref{sec:SchroEqu} that $\bar{V}_2$ can be reduced to three cases in (\ref{eq:AllPotentialCases}) depending om
$r < 0, r = 0, r > 0$, respectively, where the first order term in $\bar{V}_2$ has a form similar to $V_1$.
\medskip

\noindent
As an approximation to $\hat{V}$ we use,
\begin{equation}
\bar{V} = V_1 + \bar{V}_2.
\label{eq:VBar}
\end{equation}
The modified Schr\"odinger equation that we consider is given by  
\begin{equation}
-\frac{\sigma^2}{2\nu}\nabla^2 \Psi  + \bar{V} \Psi = E \Psi,
\label{eq:SchrodingerNew}
\end{equation}
The solution of this modified Schr\"odinger equation is derived in Section \ref{sec:SchroEqu}. The solution
is summarized in Result B.
\medskip\medskip\bigskip

\noindent
{\bf Result B}
\medskip\medskip

\noindent
{\em The explicit solution of the modified Schr\"odinger equation, more generally in a three-dimensional $P_1$-centered inertial coordinates, $(Y_1, Y_2, Y_3)$, (\ref{eq:SchrodingerNew}), is given by,
 \begin{equation}  
\Psi =  R_{\tilde{n},l}(r)Y_{m_l,l}(\phi,\theta) + \mathcal{O}(m_0m_2).
\label{eq:wave}
\end{equation}
$\tilde{n}= 0, 1,2 \ldots; l = 0, 1, 2 , \ldots; -l \leq m_l \leq l$, $r$ is the distance from $P_0$ to $P_1$, $r \geq 0$, $\phi \in [0, 2\pi]$  is the angle in the $Y_1, Y_2$-plane relative to the $Y_1$-axis, $\theta \in [0, \pi]$ is the angle relative to the $Y_3$-axis. 
$ R_{\tilde{n},l}(r)$ is given by (\ref{eq:R-detailed}) defined using Laguerre polynomials.  $Y_{l,m_l} = \Phi_{m_l}(\phi) \Theta_{l,m_l}(\theta)$ are spherical harmonic functions, where $\Phi_{m_l}(\phi), \Theta_{l,m_l}(\theta)$ are given by (\ref{eq:PhiSolution}), (\ref{eq:ThetaSolution}), respectively. $\Theta(\theta)$ is defined by Legendre polynomials.
\medskip

\noindent
$|\Psi|^2$ is the probability distribution function of finding $P_0$ at a location $(r,\phi,\theta)$. In particular, the radial probability distribution function of finding $P_0$ at a radial distance $r$ is given by   
\begin{equation}
P(r)  = R^2(r)r^2   + \mathcal{O}(m_0m_2) ,
\label{eq:Probability}
\end{equation}
where $R \equiv R_{\tilde{n},l}$.
\medskip

\noindent
$\Psi$ exists provided the energy, $E$, is quantized as,
\begin{equation}
E \equiv \hat{E}_{\tilde{n}} =  -\frac{4 \sigma}{\tilde{n}^{2}} + \mathcal{O}(m_0m_2), 
\label{eq:SchroEnergy}
\end{equation}
where the remainder term is smooth in $Y_1, Y_2, Y_3$.}
\medskip

\noindent 
\medskip\medskip\bigskip

\noindent It is noted that the solution of (\ref{eq:SchrodingerNew}) is valid for $\bar{V}=0$. However, we assume $\bar{V} \neq  0$ to compare with the resonance solutions of three-body problem, where $Y_3 = 0$.   All the terms $\mathcal{O}(m_0m_2)$  are  smooth in $Y_1,Y_2,Y_3$. 
\medskip

\noindent
It is assumed that $m_0$ can be taken sufficiently small, such that for any given small number, $\epsilon_2 > 0$, and for $(Y_1, Y_2, Y_3) \in D$, $D$ compact,  the term $ \mathcal{O}(m_0m_2)$ in (\ref{eq:SchroEnergy})  satisfies,  $|\mathcal{O}(m_0m_2)| <  \epsilon_2$.
In this sense, $ \hat{E}_{\tilde{n}}  \approx -\frac{4 \sigma}{\tilde{n}^{2}}$ .   
\medskip

\noindent
The planar case is now assumed, $Y_3=0$, unless otherwise indicated.
\medskip

\noindent
{\em Assumptions 1} \hspace{.05in} The use of the approximate symbol for $\hat{E}_{\tilde{n}}$ using $ \mathcal{O}(m_0m_2)$, is different to the one given in Definition 1, for $\omega_1$ using  $\mathcal{O}(\delta)$, $\delta=m_2/m_1$. In Definition 1, $\mathcal{O}(\delta)$ depends on $t$, and to bound it by a given small number $\epsilon_1$, $t$ varies on a compact set and $m_2$ is taken sufficiently small. In the second case, $\mathcal{O}(m_0m_2)$, depends on $(Y_1,Y_2)$, and to bound it by a small number $\epsilon_2$, $(Y_1,Y_2)$ varies on a compact set $D$ and $m_0$ is taken sufficiently small. To satisfy both cases, it is necessary to assume $m_0, m_1$ are sufficiently small. The use of $\approx$ is taken from context.
\medskip\medskip

\subsection{Equivalence of Quantized Energies with Resonance Solutions }
\medskip\medskip

The quantized values of the energy, $\hat{E}_{\tilde{n}}$, (\ref{eq:SchroEnergy}), are for the modified Schr\"odinger equation, (\ref{eq:SchrodingerNew}).  These energies are not obtained for the three-body problem, but result from an entirely different modeling. When they are substituted for $\tilde{E}_1$ in the two-body energy relation, (\ref{eq:Planck2}), for $P_0$ moving about $P_1$, it is calculated in Section \ref{sec:SchroEqu}, (\ref{eq:FrequencyCalc}), that they yield rational values for the two-body frequency, $\omega_1$,
\begin{equation}
\omega_1 |_{{\tilde{E}_1} = \hat{E}_{\tilde{n}} }   \equiv \omega_1(\tilde{n}) = \frac{8}{\tilde{n}^{3}}   +  \mathcal{O}(m_0m_2)  .
\label{eq:SpecialTwoBodyFrequencies}                                 
\end{equation}   
It is significant that the leading dominant term of $\omega_1(\tilde{n})$
is independent of masses and distances. This implies that to first order the frequencies do not depend on the masses or any other parameters. 
\medskip

\noindent
Thus, for $\tilde{n} = 1,2, \ldots $, infinitely many frequencies are obtained, $\omega_1(\tilde{n})$. We would like to show that these frequencies correspond to $\tilde{n}$ resonance orbits of $\mathfrak{F}$ for the three-body problem. Hence, we need to compare the frequencies $\omega_1(\tilde{n})$, given by (\ref{eq:SpecialTwoBodyFrequencies}), with the frequencies $\omega_1(m/n)$, defined by (\ref{eq:ResFrequenciesSec2}). The following result is obtained, 
\medskip\medskip

\noindent
{\bf Result C} 
\medskip\medskip

\noindent
{\em The quantized energy values, (\ref{eq:SchroEnergy}), of the modified Schr\"odinger equation can be put into a one to one correspondence with a subset, $\mathfrak{U}$, of the resonance solutions $\Phi_{m/n}(t) \in \mathfrak{F}$ of the circular restricted three-body problem, where
\begin{equation}
\mathfrak{U} = \{ \Phi_{m/n}(t)| m = 8, n = \tilde{n}^3, \tilde{n} = 1,2, \ldots\} .
\end{equation}
}
\medskip\medskip

\noindent
This is proven in Section \ref{sec:SchroEqu} by scaling the restricted three-body problem and using the fact that this scaling does not effect the leading order term  $\frac{8}{\tilde{n}^3}$ of $\omega_1(\tilde(n))$. 
\medskip
  
\noindent
It is noted that $m=8, n= \tilde{n}^3$ implies that in the time it takes $P_0$ to make $\tilde{n}^3$ cycles about $P_1$, $P_2$ makes $8$ cycles about $P_1$. 
\medskip\medskip

As previously noted, the limiting case of $m_2=0$ has been excluded in this paper since it is degenerate in the sense that the resonance families of solutions no longer exist. One can make a comparison with quantized two-body elliptic orbits of $P_0$ about $P_1$ with the classical Schr\"odinger equation for $m_2=0$ (see \cite{Sommerfeld:1921}, page 263), but this case does not yield the transitioning resonance solutions described in this paper. 
\medskip\medskip\medskip\medskip

\subsection{Quantum Scale}
\medskip\medskip

The results presented thus far are for mass values that are not in the quantum-scale.  Consider the family, $\mathfrak{U} \subset \mathfrak{F}$, of resonance periodic orbits for $P_0$ in the three-body problem, whose frequencies, $\omega_1(m,n)$, given by (\ref{eq:ResFrequenciesSec2}), where $m/n  \approx 8/\tilde{n}^3$. These frequencies correspond to the quantized energy values, $\hat{E}_{\tilde{n}}$,  of the modified Schr\"odinger equation.   When the masses, $m_k$, $k=0,1,2$, get smaller and smaller, along with the relative distances betwen the particles, as they approach the quantum-scale, $\omega_1$, $\omega$, increase in value as $r_1^{-1/2}, \beta^{-1/2}$ as $r_1, \beta \rightarrow 0$, respectively.  The particles remain gravitationally bound to each other. The mass of $P_0$ is negligible with respect to that of $P_1, P_2$. As the distances decrease, the motions of the particles produces a gravitational field by the circular motion of $P_1, P_2$ and the resonance motion of $P_0$. We refer to this gravitational field as a {\em resonance gravitational field}.
\medskip
  
\noindent
When the system of three particles reaches the quantum scale they take on a wave-particle duality. The differential equations for the three-body problem are no longer defined.   The previous resonance motion of the particles takes on a wave character. 
\medskip

\noindent
The three-body problem is no longer defined in the quantum scale and therefore Result A is no longer valid.  However, the modified Schr\"odinger equation is still well defined. We can now assume the three-dimensional wave solutions.  The quantized energy values are still defined, for $\tilde{n} = 1,2 \ldots$. Now, they are identified with pure wave solutions $\Psi(r,\theta)$ given in Result B. The values of $\hat{E}_{\tilde{n}}$, can be viewed as taking on wave resonance values.  This is summarized in, 
\medskip\medskip

\noindent
{\bf Result D}
\medskip\medskip
 
\noindent
{\em The resonance solutions $\Psi_{m/n}(t) \in \mathfrak{U}$ for $P_0$ for the three-body problem, which are given by the solutions $\Psi$, (\ref{eq:wave}), of the modified Schr\"odinger equation are also given by $\Psi$ when the masses are reduced to the quantum scale. This provides a quantization of the gravitational dynamics of $P_1$ for the motion of $m_0$ corresponding to the energies $\hat{E}_{\tilde{n}}$, given by (\ref{eq:SchroEnergy}).}     
\medskip\medskip\medskip\medskip

In the quantum scale, where $\sigma \rightarrow 0$ as $m_0, m_1 \rightarrow 0$, shown in Section \ref{sec:SchroEqu}, there is a transition of the resonance solutions into  wave solutions, as summarized in  Result D, using $\Psi$.  However, to make these wave solutions more physically relevant, we would like to have $\sigma(m_0, m_1, G) = \hslash$.  
\medskip

\noindent
It is shown in Section (\ref{sec:SchroEqu}), Proposition 4.1, that as $m_0, m_1 \rightarrow 0$, there exist mass values where $\sigma(m_0, m_1, G) = \hslash$. These mass values lie on an algebraic curve in $(m_0,m_1)$-space.
For these values of $m_0, m_1$, the term $-\frac{\sigma^2}{2\nu}\nabla^2 \Psi$ of the modified Schr\"odinger equation matches the same term of classial Schr\"odinger equation. In this case, only the gravitational potential is present. To make this 
accurate for atomic interaction, for example, for the motion of an electron about a nucleus of the Hydrogen atom, the gravitational potential 
needs to be replaced by the Coulomb potential. 
\medskip

\noindent
If we consider the  modified Schr\"odinger equation, it can be further altered by adding, for example, a Coulomb  potential. If 
the masses are chosen so that $\sigma = \hslash$, then one obtains a classical Newton-Schr\"odinger equation model \cite{Diosi:1984}, \cite{Penrose:1996}.  This could also be studied with $\sigma \neq \hslash$. 
\medskip

\noindent
The wave solutions of the modified Schr\"odinger equation could be considered in the quantum scale where $\sigma \neq \hslash$. This is not studied in this paper.  
\medskip\medskip\medskip

\section{Weak Capture and Resonant Motions in the Three-Body Problem}
\label{sec:WeakCapture} 
\medskip\medskip

In this section we show how to prove Result A. The idea of the proof of Result A is to utilize the geometry of the phase space about $P_1$, $P_2$, where the motion of $P_0$ is constrained by Hill's regions. Within the Hill's regions, the dynamics associated to weak capture from near $P_2$ together with the global properties of the invariant hyperbolic manifolds around $P_1$ will yield the proof. 
\medskip\medskip\medskip

The planar circular restricted three-body problem in inertial coordinates is defined in Section \ref{sec:Results} by (\ref{eq:DE3Body}) for the motion of $P_0$. If $P_0$ moves about $P_1$ with elliptic initial conditions, and a rotating coordinate system is assumed that rotates with the same constant circular frequency $\omega$ between $P_1$ and $P_2$, then the motion is understood by the Kolmogorov-Arnold-Moser(KAM)Theorem \cite{SiegelMoser:1971},\cite{Belbruno:2004}.  It says that nearly all initial elliptic initial conditions of $P_0$ with respect to $P_1$ give rise to quasi-periodic motion, of the two frequencies, $\omega_1, \omega$, where $\omega_1$ is the frequency of the elliptic motion of $P_0$ about $P_1$, provided they satisfy the condition that $\omega_1/\omega$ is sufficiently non-rational. For the relatively small set of motions of $P_0$ where $\omega_1/\omega$ is sufficiently close to a rational number, the motion is chaotic. It is also necessary to assume that $\mu = m_2/(m_1+m_2)$ is sufficiently small. 

The planar modeling is assumed without loss of generality. This follows since the resonance orbits we will be considering for $P_0$ moving about $P_1$ are approximately two-body in nature. This implies approximate planar motion. These same orbits result from weak capture conditions and escape, which imply that the plane of motion of $P_0$ about $P_1$ will approximately be the same plane of motion as that of $P_2$ about $P_1$.  Thus, co-planar modeling assumed in the restricted three-body problem is a reasonable assumption. 

Whereas the motion of $P_0$ about $P_1$ is well understood by the KAM theorem for small $m_2$, the general motion of $P_0$ about $P_2$ is not well understood since it's considerably more unstable. The instability arises due to the fact that $m_2$ is much smaller than $m_1$, and the KAM theorm cannot be easily applied unless $P_0$ moves infinitely close to $P_2$ \cite{Kummer:1979}. This implies that if  $P_0$ starts with an initial two-body elliptic state with respect to $P_2$,  its trajectory is substantially perturbed by the gravitational effect of $P_1$. The resulting motion of $P_0$ about $P_2$ is unstable and generally rapidly deviates from the initial elliptic state. Numerical simulations show the motion to be chaotic in nature. Results described in this section provide a way to better understand motion about $P_2$. 

The notion of weak capture (defined in Section \ref{sec:Results})  of $P_0$ about $P_2$ is useful in trying to 
understand the motion of $P_0$ about $P_2$ with initial elliptic conditions.  The idea is to numerically propagate trajectories of the three-body problem with initial conditions that have negative energy, $E_2 < 0$, with respect to $P_2$, and measure how they cycle about $P_2$, described in more detail later in this section. Generally, if $P_0$ performs $k$ complete cycles about $P_2$, relative to a reference line emanating from $P_2$, without cycling about $P_1$,  then the motion of $P_2$ is called 'stable',  provided it returns to the line with $E_2 < 0$, while if does not return to the line after $(k-1)$ complete cycles, and cycles about $P_1$,
the motion is called 'unstable'. It is also called unstable if $P_0$ does return to the line, but where $E_2 > 0$.  (see Figure \ref{fig:Fig180})  The line represents a two-dimensional surface of section in the four-dimensional phase space, $\mathbb{R}^4$.
The set of all stable points about $P_2$ defines the '$k$th stable set', $W^s_k$, and the set of all unstable points is called the '$k$th unstable set', $W^u_k$. Points that lie on the boundary between $W^s_k$ and $W^u_k$ define a set, $\mathcal{W}_k$, called the '$k$th weak stability boundary'.  The boundary points are determined algorithmically, by iterating between stable and unstable points \cite{BGT:2010}.

Points that belong to $W^u_k$ are in weak capture with respect to $P_2$ since they start with $E_2 < 0$,  which lead to escape with $E_2 > 0$ (Proposition 3.1).  However, this may not be the case for points in $W^s_k$ since after they cycle about $P_2$ $k$ times, it is possible they can remain moving 
about $P_2$ for all future time and  $E_2$ will be negative each time $P_0$ intersects the line.

$\mathcal{W}_k$ was first defined in \cite{Belbruno:1987}, for the case $k=1$. This set has proved to have important applications in astrodynamics to enable spacecraft to transfer to the Moon and automatically go into weak capture about the Moon, that requires no fuel for capture. This was a substantial improvement to the Hohmann transfer, which requires substantial fuel for capture \cite{BelbrunoMiller:1993}, \cite{Belbruno:2004}. %\cite{Belbruno:2007}   
\footnote{It was first used operationally in 1991 to rescue a Japanese lunar mission by providing a new type of transfer from the Earth to the Moon used by its spacecraft, {\em Hiten}. } It also has applications in astrophysics on the Lithopanspermia Hypothesis \cite{BMMS:2012}. The weak stability boundary was generalized to $k$-cycles, $k > 1$, with new details about its geometric structure in \cite{GarciaGomez:2007}. \cite{BGT:2010} makes an equivalence of  $\mathcal{W}_k$ with the stable manifolds of the Lyapunov orbits associated to collinear Lagrange points.
\medskip\medskip

\noindent
$W^s_k$, $W^u_k$, $\mathcal{W}_k$ are defined more precisely: 
\medskip\medskip

\noindent
We transform from $X = (X_1, X_2)$ defined in (\ref{eq:DE3Body}) to a rotating coordinate system, $x = (x_1, x_2)$, that rotates with frequency $\omega$, so that in this system, $P_1, P_2$ are fixed on the $x_1$-axis.  Scaling $m_1 = 1 -\mu, m_2 = \mu, \mu > 0, G=1 , \beta= 1, \omega = 1$, as mentioned in Section \ref{sec:Results}, we place $P_1$ at $x = (\mu,0)$ and $P_2$ at $(-1+\mu,0)$.  (\ref{eq:DE3Body}) becomes,
\begin{equation}
\ddot{x} +2(-\dot{x}_2, \dot{x}_1)= \tilde{\Omega}_{x}  , 
\label{eq:DE3BodyRot}
\end{equation}
$\tilde{\Omega} = (1/2)|x|^2 + (1-\mu)r_1^{-1} + \mu r_2^{-1} +(1/2)\mu(1-\mu)$ ,
$r_1 = |(x- (\mu,0)|$, $r_2 = |x - (-1+\mu,0)|$. The Jacobi integral function, $J(x, \dot{x})$ for this system is given by
\begin{equation}
J = 2\tilde{\Omega} - |\dot{x}|^2. 
\label{eq:Jacobi}
\end{equation}
The differential equations have $5$ well known equlibrium points, $L_i$, $i=1,2,3,4,5$, where $L_1, L_2,L_3$ are the collinear Lagrange points, and $L_4,L_5$ are equilateral points. We assume the convention that $L_2$ lies between $P_1, P_2$. $J|_{L_i} = C_i$, where $3=C_4=C_5< C_3< C_1 < C_2$.  The collinear points are all local saddle-center points with eigenvalues, $\pm \alpha$ and $\pm i \beta$, $\alpha > 0, \beta > 0, i^2 = -1$. The equilateral points $L_4, L_5$ are locally elliptic points.  We will focus on $L_1, L_2$ in our analysis. As is described in \cite{Belbruno:2004}, \cite{LMS:1985}, the distance of $L_1, L_2$ to $P_2$, $r_{L_j}$, $j=1,2$, is $r_{L_j} = \mathcal{O}(\mu^{1/3})$. $C_j = 3 + |\mathcal{O}(\mu^{2/3})| \gtrsim 3$ for $\mu \gtrsim 0$. 
\medskip

\noindent Projecting the three-dimensional Jacobi surface $J^{-1}(C)$ into physical $(x_1, x_2)$-space, yields the Hill's regions,  where $P_0$ is constrained to move. (see \cite{Belbruno:2004}, Figure 3.6) For $C$ slightly greater than $C_2$, $C \gtrsim C_2$, the Hill's regions about $P_1, P_2$, labeled $H_1,H_2$, respectively, are not connected, so that $P_0$ cannot pass from one region to another. There is also a third Hill's region, $H_3$ that surrounds both $P_1, P_2$ disconnected from $H_1, H_2$, where $P_0$ can move about both primaries. When $C=C_2$, $H_1, H_2$ are connected at the single Lagrange point $L_2$ and $P_0$ still cannot pass between the primaries. When $C \lesssim C_2$, a small opening occurs between $P_1, P_2$ near the $L_2$ location, we refer to as a neck region, $N_2$, first discussed in \cite{Conley:1968}. When $C$ decreases further, $C \lesssim C_1$, another opening occurs near 
$L_1$ and forms another neck region, $N_1$, that connects $H_2$ with the outer Hill's region, $H_3$.
\medskip

\medskip\medskip\medskip

\begin{figure}
\centering	\includegraphics[width=0.70\textwidth, clip, keepaspectratio]{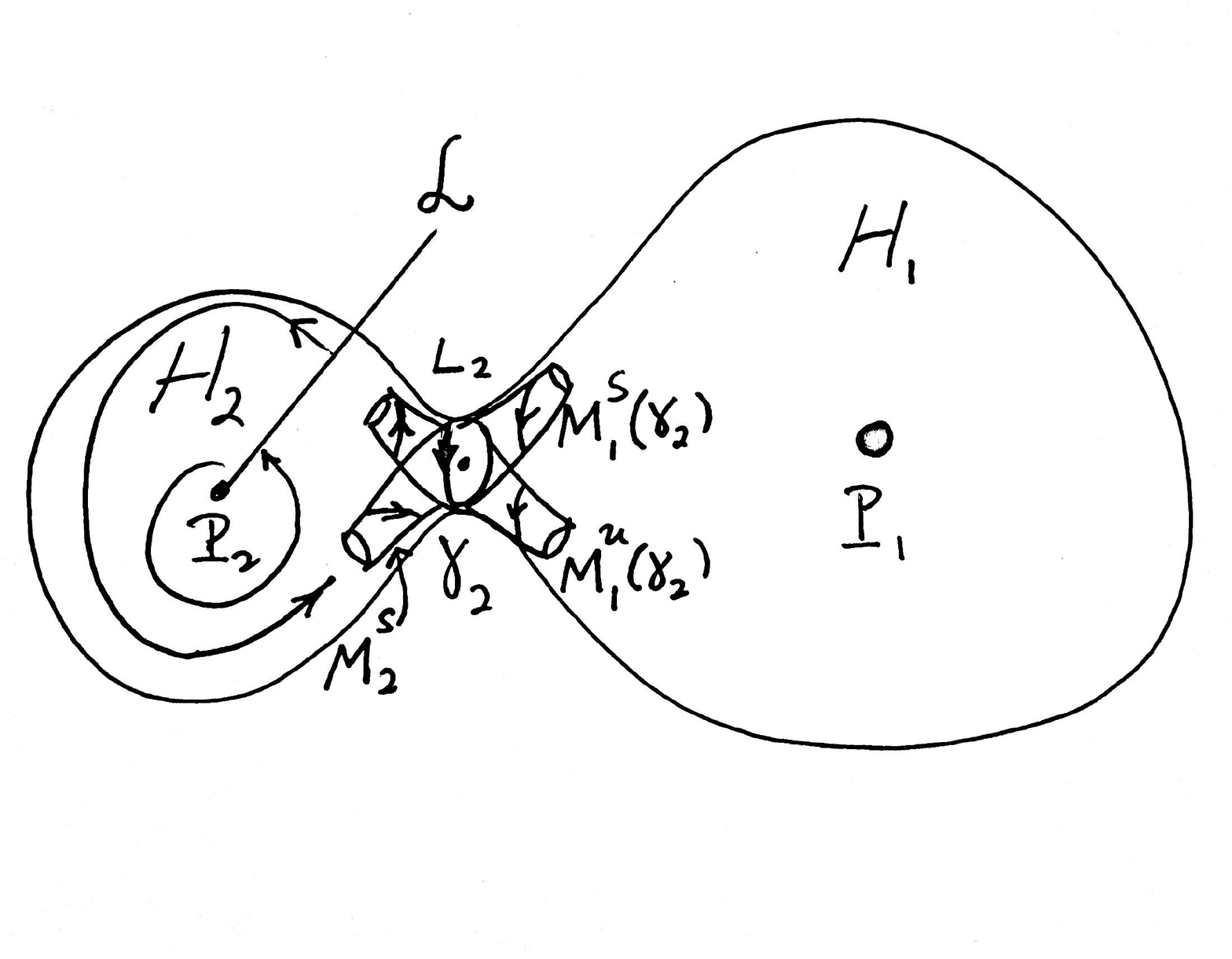}
\caption{ Hill\rq{}s regions, $H_1, H_2$ connected in neck region about  unstable Lyapunov orbit $\gamma_2$, $C \lesssim C_2$.   Cylinderical stable and unstable manifolds, $M^{s,u}_k$, $k=1,2$, in respective regions $H_k$, shown projected from four-dimensional position-velocity space into position space. The dot inside  $\gamma_2$ is the location of $L_2$ when $C=C_2$ . $P_0$ can only move from $H_2$ to $H_1$ through $M^s_2$ and then $M^u_1$ from an unstable point on $\mathcal{L}$. This illustrates the separatrix property of the manifolds.  (This projection is not to scale and shows $M_2^s, M_1^u$, that exist in 4-dimensional phase space,  projected into physical space.  It is meant to give an idea of the geometry)    }
\label{fig:Fig140}
\end{figure}

\noindent
A retrograde unstable hyperbolic periodic orbit is contained in $N_2$, we label $\gamma_2$. $\gamma_2$ has local stable and unstable two- dimensional manifolds $M^s_j(\gamma_2), M^u_j(\gamma_2)$, $j=1,2$, which extend from $N_2$ into $H_j$. These manifolds are topologically equivalent to two-dimensional cylinders.  It is shown in \cite{Conley:1968} that orbits can only pass from $H_2$ to $H_1$, or from $H_1$ to $H_2$, by passing within the three-dimensional region contained inside $M^s_j(\gamma_2), M^u_j(\gamma_2)$, which are called transit orbits. For example, to pass from $H_2$ to $H_1$, $P_0$ must pass into the three-dimensional region inside $M^s_2(\gamma_2) \subset H_2$ and out from the region inside $M^u_1(\gamma_2) \subset H_1$  (see Figure \ref{fig:Fig140})    (see also \cite{Belbruno:2004}, Figure 3.9). $N_2$ is bounded on either side of $P_2$ by vertical lines $l_R, l_L$, that cut the $x_1$-axis, to the right and left of $P_2$, respectively. On the Jacobi surface, $\{ J =C\}$, $J^{-1}(N_2)$ is a set with topological two-dimensional spheres as boundaries, $S^2_R, S^2_L$ corresponding to the lift of $l_R, l_L$, respectively, onto $\{ J =C\}$.  When a transit oribit passes from $H_2$ to $H_1$, then on the Jacobi surface, $P_0$ passes from $ S^2_L$ to $S^2_R$.  The bounding spheres separate $H_1, H_2$ from $N_2$. 
\medskip

\begin{figure}
\centering
	\includegraphics[width=0.60\textwidth, clip, keepaspectratio]{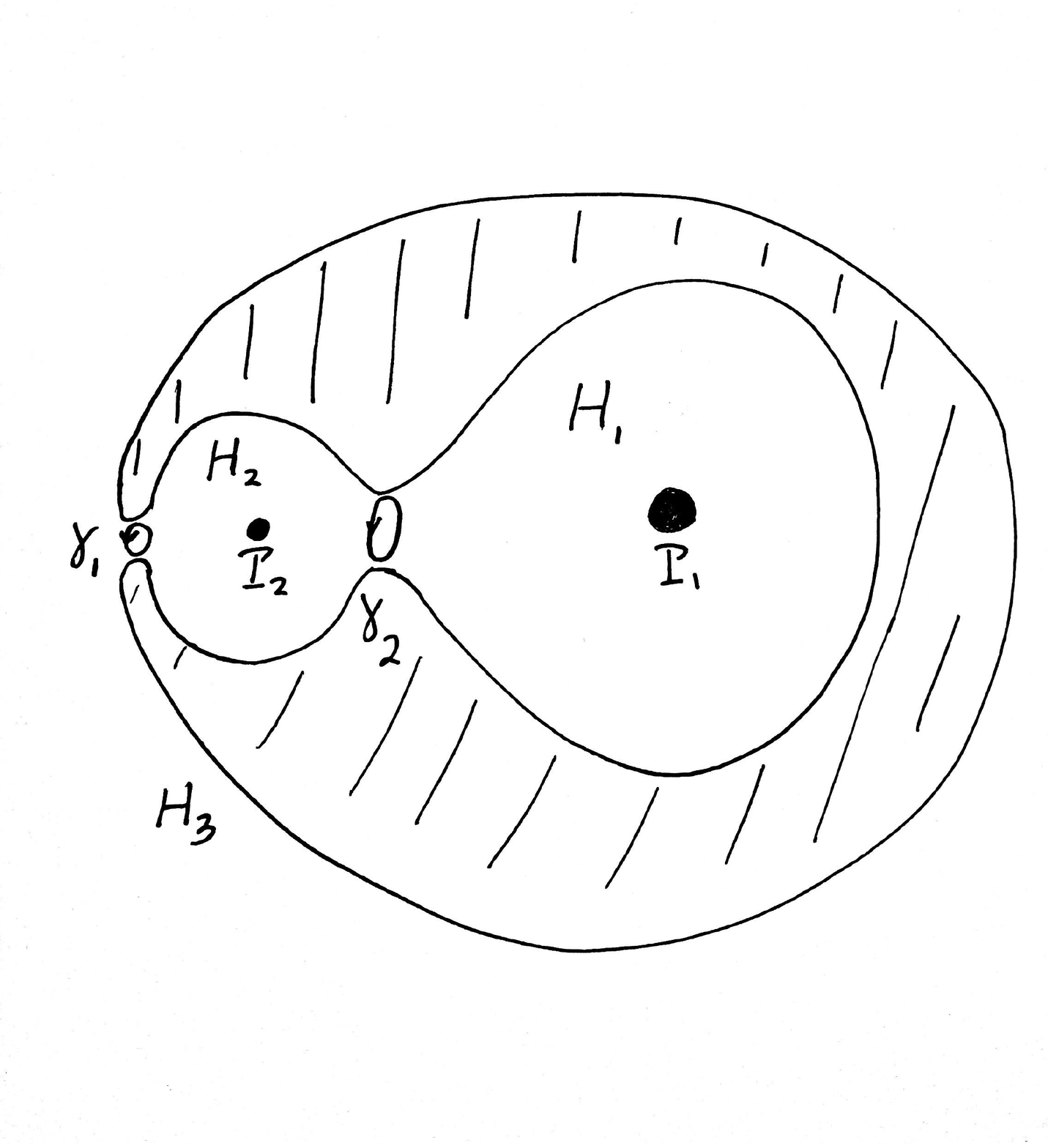}
\caption{ Hill\rq{}s regions, $H_1, H_2$ connected in neck region about  unstable Lyapunov orbit $\gamma_2$, and also about $\gamma_1$,  for $C \lesssim C_1$. The opening about $\gamma_1$ connects $H_2$ to a larger Hill\rq{}s region, $H_3$,  about $H_1, H_2$. The hatched region is where the point $P_0$ cannot move for the given Jacobi energy $C$. (This is a sketch and  not to scale. It is meant to give an idea of the geometry)   }
\label{fig:Fig150}
\end{figure}

\begin{figure}
\centering
	\includegraphics[width=0.70\textwidth, clip, keepaspectratio]{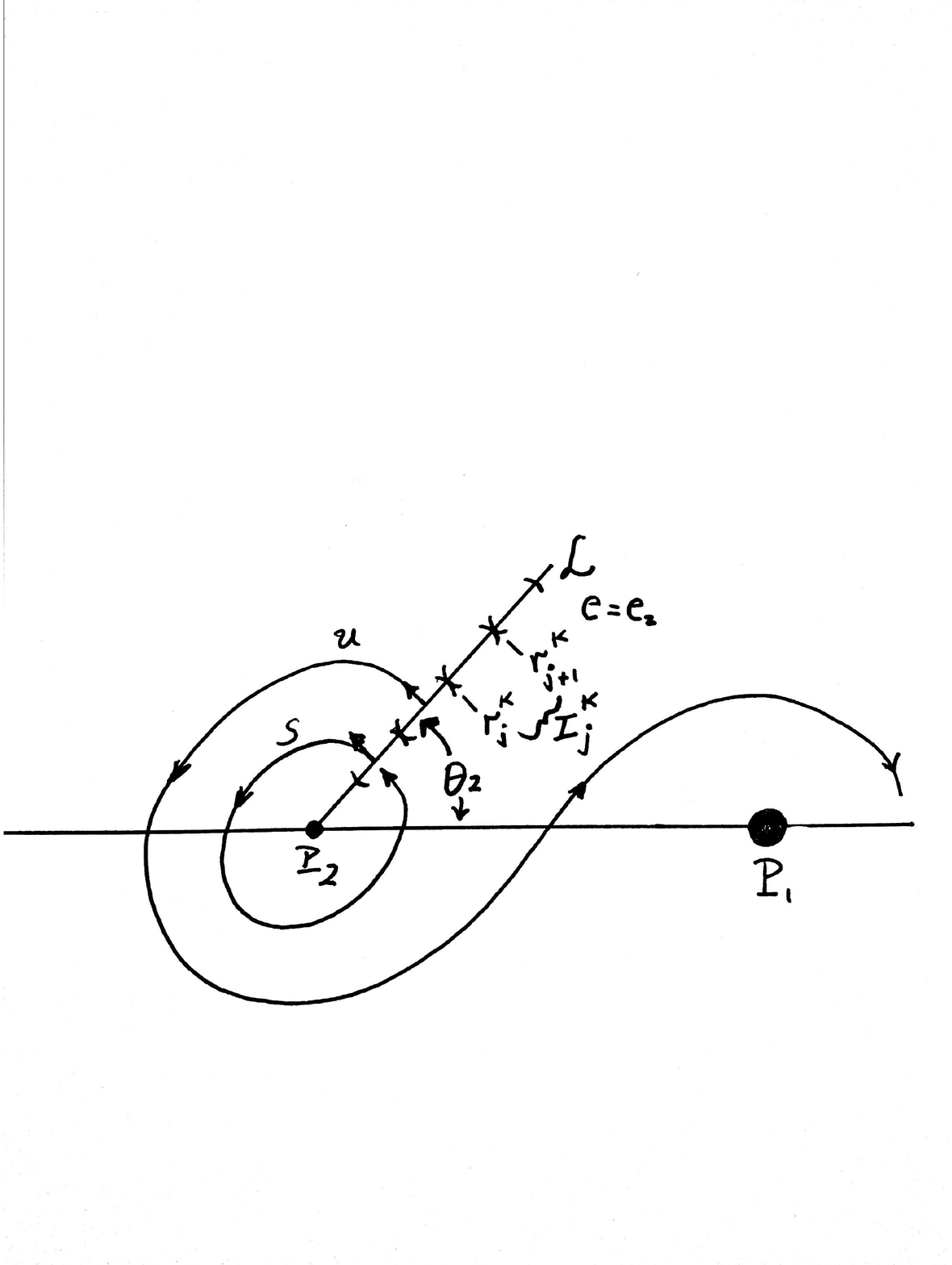}
\caption{ Alternating stable and unstable intervals $I^k_j$ on $\mathcal{L}$ for $k=1$ cycles illustrating the algorithm. The boundary points $r_j^k$ belong to the weak stability boundary. Unstable points in the intervals leading to unstable motion(labeled $u$). These points are weakly captured. Stable motion, labled $s$, for points in stable intervals.  (Sketch, not to scale)}   
\label{fig:Fig180}
\end{figure}

\noindent
For $C \lesssim C_1$, $N_1$ contains the Lyapunov orbit $\gamma_1$. Manifolds,  $M^s_j(\gamma_2), M^u_j(\gamma_2)$, $j=2,3$, are similarly obtained where transit orbits can pass between $H_2$ and $H_3$, passing through the respective bounding spheres. The geometry in this case is shown in Figure \ref{fig:Fig150}.)
\medskip

\medskip
It is noted that in center of mass, rotating coordinates, $(x_1, x_2$), (\ref{eq:KepEng}) becomes,
\begin{equation}
E_{2R} =  {1\over2} |\dot{{x}}|^2 + {1\over2} |{x}|^2 - \omega(\dot{x}_1x_2 - \dot{x}_2x_1)   - {G(m_0 + m_2)\over r_2},
\label{eq:KepEngRotate}
\end{equation}
with $\omega = 1, m_0 = 0, m_2 = \mu, G=1$.  
\medskip

\noindent
Finally, a translation, $z_1 = x_1 - (-1+\mu), z_2 = x_2$,  is made to a $P_2$-centered coordinate system, $(z_1, z_2)$, where  $P_1$ is at $(1,0)$, and $r_2 = |z|$.  For notation, we set $r \equiv r_2$.  We refer to $E_{2R}$ in center of mass rotating coordinates $(x_1, x_2)$ and also in $P_2$-centered rotating coordinates $(z_1, z_1)$, where $E_{2R}$ is a different expression from (\ref{eq:KepEngRotate}). 
\medskip

\noindent
The line $\mathcal{L}$ emanates from $P_2$ and makes an angle $\theta_2 \in [0, 2\pi]$ with respect to the $z_1$-axis. Trajectories of $P_0$ are propagated from $\mathcal{L}$ such that at each point on $\mathcal{L}$ at a distance $r > 0$, the eccentricity, $e_2$, is kept fixed to a value, $e_2 \in [0, 1)$, by adjusting the velocity magnitude, whose initial direction is perpendicular to $\mathcal{L}$. Also, the velocity direction is assumed to be clockwise (similar results are obtained for counter clockwise propogation). The initial points of propagation on $\mathcal{L}$ are periapsis points of an osculating ellipse of velocity $v_p = \sqrt{G(m_0+m_2)(1+e_2)/r} -\omega r$, $\omega =1, m_0 = 0,m_2 = \mu$. It is noted that $\mathcal{L}$  makes a two-dimensional surface of section, defined in polar coodinates,  $S_{\theta_2^0} = \{ (r_2, \theta_2, \dot{r}_2, \dot{\theta}_2| \theta_2 = \theta_2^0, \dot{\theta}_2 > 0 \}$. It is also noted that as $r$ changes on $\mathcal{L}$, the Jacobi energy also changes. This implies that $W^u_k, W^s_k, \mathcal{W}$ do not lie on a fixed Jacobi surface. Also the Hill's regions vary within these sets.
\medskip

As is described in \cite{GarciaGomez:2007} \cite{BGT:2010}, a sequence of consecutive open intervals, $I^k_j$, are obtained along $\mathcal{L}$, for a fixed $\theta_2, e_2$, that alternate between stable and unstable points, for $k$ cycles. That is,  $I^k_1 = \{r^k_0 < r < r^k_1\}$, $r^k_0=0$, are stable points, $I^k_2 = \{ r^k_1 < r <r^k_2 \}$ are unstable points, etc.   (see Figure \ref{fig:Fig180})
  There are $N_k(\theta_2, e_2)$ stable sets, and unstable sets, for an integer $N_k \geq  1$. 
The boundary points $r^k_j, j = 1,2, \ldots, N_k(\theta_2, e_2)$, represent the transition between stable and unstable points relative to $k$ cycles, where the $k$th unstable points lead to stable motion for $k-1$ cycles and are unstable on the $k$th cycle. The $kth$ stable set for a given value of $\theta_2, e_2$  is given by,  
\begin{equation}
W^s_k(\theta_2, \e_2)  = \bigcup_{j = 0}^{N_k(\theta_2, e_2)}(r^k_{2j}, r^k_{2j+1}).
\label{eq:StableSetSection}
\end{equation}
This is a slice of the entire stable set, $W^s_k$, by varying $\theta_2, e_2$, given by
\begin{equation}
W^s_k  = \bigcup_{\theta \in [0,2\pi], e \in[0,1)} W^s_k(\theta_2, \e_2).
\label{eq:StableSet}
\end{equation}
\medskip

\noindent
We define $\mathcal{W}_k = \partial W^s_k$. $\mathcal{W}_k$ has a Cantor-like structure as is described in \cite{BGT:2010}.  
The numerical estimation of $W^s_k$, $\mathcal{W}_k$ is given in  \cite{GarciaGomez:2007},\cite{BGT:2010}, \cite{TB:2009}, for different values of $k$, $\mu$, $\theta_2$, $e_2$.  The motion of $P_0$ is seen to be unstable and sensitive for initial conditions near $\mathcal{W}_k$. It is remarked that due to limitations of computer processing time, $k$ is not taken too large. 
 \medskip

\noindent
A main result of \cite{BGT:2010}, is that $\mathcal{W}_k$ about $P_2$ is equivalent to the set of global stable manifolds, $M_2^s(\gamma_1) \cup M_2^s(\gamma_2)$, to the Lyapunov orbits, $\gamma_1, \gamma_2$, respectively, about the collinear Lagrange points, $L_1, L_2$, on either side of $P_2$ for $C \lesssim C_1$ and $\mu$ sufficiently small, in $H_2$. Similarly, one could restrict $C \lesssim C_2$ and have equivalence to only the global manifold  $M_2^s(\gamma_2)$ in $H_2$.  This is demonstrated numerically by examining the intersections of $M_2^s(\gamma_1), M_2^s(\gamma_2)$ on surfaces of section $S_{\theta_2^0}$ satisfying $\dot{r} = 0, E_2 < 0$ for $\mu$ sufficiently small, and varying $0 \leq \theta_2^0 \leq 2\pi$.  
It is shown in \cite{BGT:2010} that a very small set of points exist on $\mathcal{W}_k$ that do not satisfy this equivalence. These points are not considered. \footnote{It is noted that the proof of equivalence of $\mathcal{W}_k$ with the global stable manifolds to $L_1, L_2$ in \cite{BGT:2010} is numerically supported and based on rigorous analytical estimates. Thus, the proof is rigorous in that sense. This is also true of the structure of $\mathcal{W}_k$  obtained in \cite{GarciaGomez:2007}. A purely analytic proof for the global manifold  stricture about $P_2$ and  $\mathcal{W}_k$  is not available at this time.  However, in the case of motion about $P_1$, the analogous structure of  $\mathcal{W}_k$ is analytically proven \cite{BGT:2013}.}
\medskip

\noindent  The reason this equivalence is true is due to the separatrix property of the manifolds (see \cite{Conley:1968} \cite{BGT:2010}). Assume $C \lesssim C_2$. The separatrix property means that if a trajectory point is inside of the region bounded by  $M_2^s(\gamma_2)$ on $\mathcal{L}$, it will wind about $P_2$ staying inside the region contained by $M_2^s(\gamma_2)$ as $M_2^s(\gamma_2)$ winds about $P_2$. $P_0$ can't go outside this manifold region.  Eventually, $M_2^s(\gamma_2)$  will go to $\gamma_2$, and $P_0$ will pass through $N_2$ into $H_1$ as a transit orbit, after it makes $k-1$ complete cycles, before completing the $k$th cycle. This corresponds to an unstable point on the set $W^u_k$. If a trajectory point is outside  $M_2^s(\gamma_2)$ on $\mathcal{L}$ , $P_0$ will remain in $H_2$, making $k$ complete cycles about $P_2$ near the outside of  $M_2^s(\gamma_2)$, but it can't escape to $H_1$. Thus, $M_2^s(\gamma_2)$ itself is equivalent to $\mathcal{W}_k$. That is, $M_2^s(\gamma_2)$ separates between stable and unstable motion. 
\medskip

\noindent
The intersections of $M_2^s(\gamma_2)$ on $\mathcal{L}$ in physical space as it cycles around $P_2$ give rise to the alternating intervals between stable and unstable motion, $\{I^k_1, I^k_2, \ldots \}$, where there are $N_k(\theta_2, e_2)$ such intervals. 
Points inside $M_2^s(\gamma_2)$ on $\mathcal{L}$ correspond to points in the set $W_k^u$, and points outside of $M_2^s(\gamma_2)$ on $\mathcal{L}$, and close to it,  correspond to points of the set $W_k^s$.
\medskip

\begin{figure}
\centering
	\includegraphics[width=0.70\textwidth, clip, keepaspectratio]{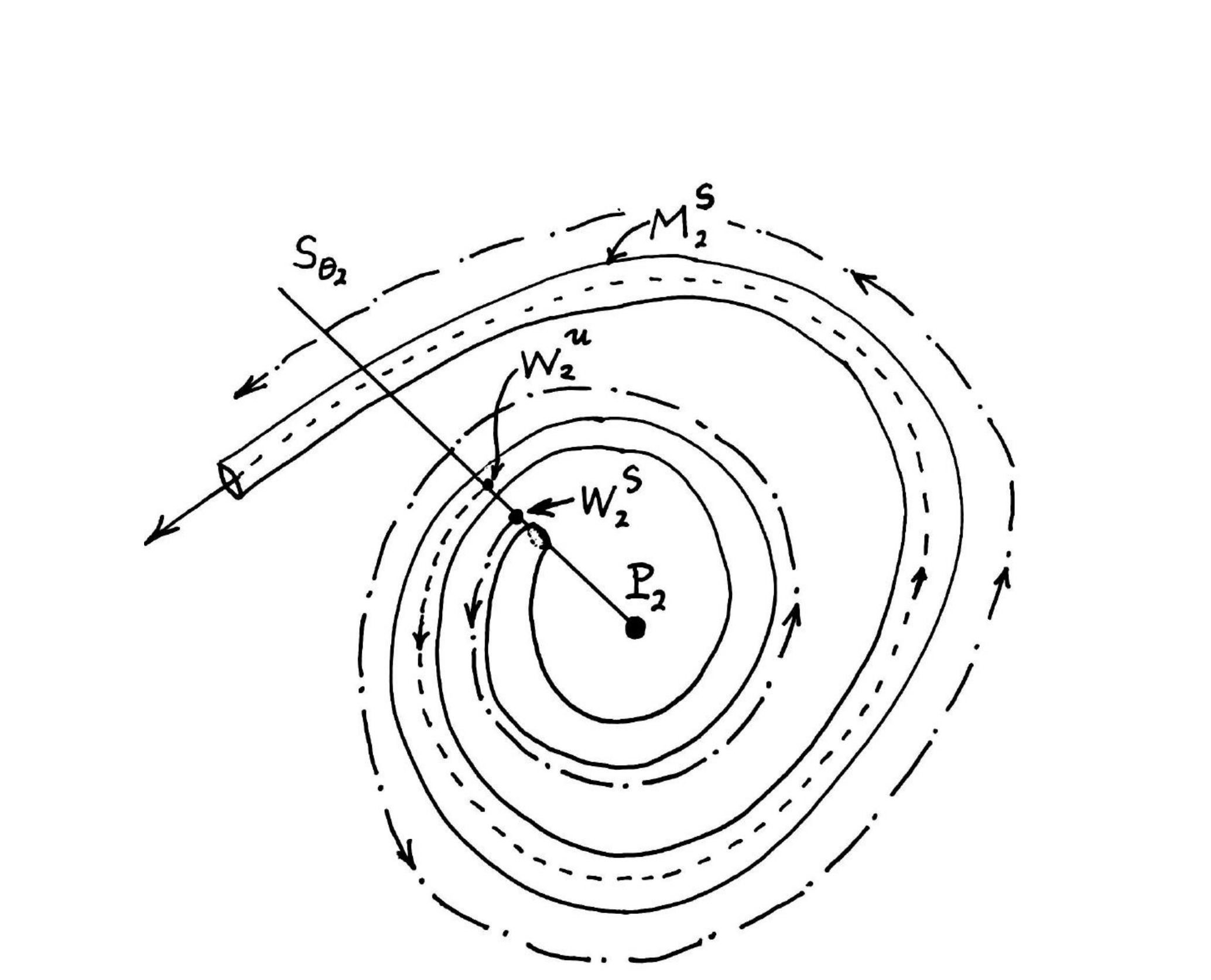}
%\centerline{\includegraphics{Fig1}}
\caption{The relationship between the manifolds, $M_2^s(\gamma_2), M_2^u(\gamma_2)$, projected into physical space and the stable and unstable sets, $W^s_k, W^u_k$,  relative to $k=2$ cycles about $P_2$. The stable points make two complete cycles about $P_2$ in $H_2$, while the unstable points transition to the $H_1$ region after the first cycle. One sees alternating stable and unstable intervals on the section $S_{\theta_2}$. (This projection is not accurate and not to scale. It shows $M_2^s$ projected onto physical space. It is meant to give a rough idea of the geometry) } 
\label{fig:Fig2}
\end{figure}

\noindent
The relationship between the manifolds and $W_k^s, W_k^u$ is shown in Figure \ref{fig:Fig2}.
\medskip
\medskip

\noindent
 It can be shown that if $M_2^s(\gamma_2)$ has transverse intersections, which is numerically demonstrated, where the manifold tube breaks, the separatrix property is still satisfied even though a section through the tube no longer gives a circle, but rather parts of broken up circles.  
\medskip

\noindent 
Numerical simulations in \cite{LMS:1985}, \cite{BGT:2010}, indicate for $C \lesssim C_1$, that $M_2^s(\gamma_1)$ can intersect $M_2^u(\gamma_2)$ transversally, giving rise to a complex network of invariant manifolds about $P_2$, and for $C \lesssim C_2$, $M_2^u(\gamma_2)$ can intersect $M_2^s(\gamma_2)$ transversally, for a set of $\mu$ and $C$. This supports the fact that the motion near \textbf{$\mathcal{W}_k$} is sensitive.  
\medskip\medskip\medskip\medskip

Let $\zeta(t) = (z_1(t), z_2(t), \dot{z}_1(t), \dot{z}_2(t)) \in \mathbb{R}^4$ be the trajectory of $P_0$ in rotating $P_2$-centered coordinates, and  $z(t) = (z_1(t), z_2(t)) \in \mathbb{R}^2$ the trajectory of $P_0$ in physical coordinates. Similarly, in inertial $P_2$ centered coordinates, we define, $\mathcal{Z}(t) = (Z_1(t), Z_2(t), \dot{Z}_1(t), \dot{Z}_2(t)$, and  $Z(t) = (Z_1(t), Z_2(t))$. We will use inertial and rotating coordinates to describe the motion of $P_0$, 
\medskip\medskip

The following result, referenced previously, is proven,  
\medskip\medskip

\noindent
{\em Proposition 3.1} ($P_0 \in W^u_k$ implies $P_0$ is weakly captured by $P_2$)
\medskip

\noindent
{\em Assume $P_0 \in W^u_k$ at time $t=t_0$, which implies $E_2(t_0) < 0$. There are two possibilities: (i.)  $P_0$ cycles about $P_2$ $k-1$ times, then moves to cycle about $P_1$ without cycling about $P_2$. This implies that $P_0$ weakly escapes $P_2$. That is, there exists a time $t^* > t_0$, after the $(k-1)$st cycle where $E_2(t^*) =0$, $E_2 \gtrsim 0$ for $t \gtrsim t^*$, and $E_2(t) < 0$ for $t_0 \leq t < t^*$, (ii.) $P_0$ does $k$ complete cycles about $P_2$, where on the $k$th cycle $P_0$ returns to $\mathcal{L}$ with $E_2 > 0$.  (It is assumed the set of collision orbits to $P_1$ and $P_2$, $\Gamma$, are excluded which are a set of measure $0$.)   }
\medskip\medskip                              
                          
\noindent 
Proof of Proposition 3.1 - (i.) This is shown to be true by noting that when $P_0$ does a cycle about $P_1$, it will cross the $x_1$-axis, where $r_2 > 1+c$, $c > 0$. (\ref{eq:KepEngRotate}) implies that $E_{2R} = (1/2)\dot{x}_1^2 + (1/2)\mu^2 
+ \dot{x}_2 (1+c) - \mu(1+c)^{-1}$, where $\dot{x_2} > 0$, $\mu \ll 1$. This implies there exists a time $t^{**}$ where $E_{2R}>0$. Since $E_{2R} < 0$ at $t = t_0$, then there exists a time $t_0 < t^* < t^{**}$ where $E_{2R}=0$ and $E_2 \gtrsim 0$ for $t \gtrsim t^*$.  (ii.) This yields weak capture since $E_2<0$ for $t = t_0$ and $E_2$ becomes positive. Thus, there exists a time $t^*$ where $E_2(t^*) = 0$, then becomes slightly positive.
\medskip\medskip

\noindent
{\em Global trajectory after weak capture }
\medskip\medskip

We rigorously prove Result A, that after weak capture with respect to $P_2$, $P_0$ can move onto a resonance orbit about $P_1$ in resonance with $P_2$, and then return to weak capture.  This is done by a series of Propositions.
\medskip\medskip\medskip

The following sets are defined for trajectories for $P_0$ starting in weak capture at $t = t_0$ that go to weak escape at a time $t_1 > t_0$.
\medskip\medskip

\noindent
{\em Assumptions A}
\medskip

\noindent
 Type I =  $\{ \zeta = (z, \dot{z})$ at $t_0$ is on or near $W^u_k$ ($E_2 < 0$,  $\dot{r} = 0$ or $|\dot{r}| \gtrsim 0$, resp.) $\}$
\medskip

\noindent Type II = $\{ \zeta(t_0) = (z(t_0), \dot{z}(t_0))$  is not near $W^u_k$, where $|\dot{r}(t_0)|$ is not near $0$ and $E_2(t_0) <0$ $\}$
\medskip

\noindent Type IIa = $\{ \zeta(t_0)$ is a Type II point  where $\zeta(t)$ goes to weak escape at $t_1 > t_0$ with $|\dot{r}(t_1)|$ not near $0$ $\}$  (i.e. there is no cycling about $P_2$.)
\medskip

\noindent Type IIb  = $\{  \zeta(t_0)$ is a Type II point where there exists a time $\hat{t} > t_0$, such that $\zeta(\hat{t})$ on or near $W^u_{k'}$, for some integer $k' \ge 1$ $\}$
\medskip

\noindent Case A  = $\{$ $P_0$ cycles about $P_2$ $(k-1)$ times, then moves to cycle about $P_1$ $\}$ 
\medskip

\noindent Case B = $\{$  $P_0$ does not cycle about $P_1$ after $(k-1)$st cycle. Instead, on $k$-th cycle about $P_2$, $P_0$ returns to $\mathcal{L}$ with $E_2 > 0$ $\}$
\medskip

\noindent $\Gamma$ = $\{$ $P_0$ goes to collision with $P_1$ or $P_2$ for $t > t_0$ $\}$ 
\medskip\medskip

\noindent Result A is stated more precisely as,
\medskip\medskip

\noindent
{\em Theorem A} \hspace{.05in} {\em Assume $P_0$ is weakly captured at a distance $r$ from $P_2$ at $t=t_0$. Assume the weak capture point,
$\zeta(t_0)$ is of Type I, Type IIb, Case A, which are numerically observed to be generic \cite{BGT:2010}, and assume the following sets are ruled out: Type IIa, Case B, Gamma (numerically observed to be small \cite{BGT:2010}). Assume also that 
$C \lesssim C_2$, $\mu$ sufficintly small. Then $P_0$ will escape $P_2$ through $N_2$  by 
passing within the region contained within $M^s_2(\gamma_2) \subset H_2$, and moving into the $H_1$ through the region within $M^u_1(\gamma_2)$. This escape is approximately parabolic since $E_2 \approx 0$ on $S^2_R \subset H_1$. (Parabolic escape is when there exists a time $t > t_0$ where $E_2 =0$.) 
$\zeta(t)$ evolves into an approximate resonance orbit about $P_1$ with an apoapsis near $S^2_R$ of $N_2$,  peforming several cycles about $P_1$, then returns to $S^2_R$  passing through $N_2$ within the region contained within $M^s_1(\gamma_2)$ and exiting $N_2$ through the interior of $M^u_2(\gamma_2)$ and moving onto weak capture about $P_2$. The process repeats unless $P_0$ moves on any of the sets: Type IIa, Case B, $\Gamma$, or $P_0$ escapes the $P_1,P_2$-system.   
\medskip

\noindent If $C \lesssim C_1$, then $P_0$ can parabolically escape $P_2$ through $S^2_R$ of $N_2$, as previously described obtaining a sequence of resonance orbits in $H_1$, or it can parabolically escape $P_2$ through $S^2_L = \partial N_1$ into $H_3$, by passing through the region within $M_2^s(\gamma_1)$ and exiting from the region within $M_3^u(\gamma_1)$,
and form a larger resonance orbit about $P_1$ with a periapsis near $S^2_L$ in $H_3$, which eventually returns to weak capture about $P_2$ though $N_1$, reversing the previous pathway.  This process terminates if $P_0$ moves on any of the sets Type IIa, Case B, $\Gamma$ or escapes the $P_1, P_2$ system. 
\medskip

\noindent This yields a sequence of approximate resonance orbits depending on the choice of the weak capture initial condition. The set of all such resonance orbits form the family, $\mathfrak{F}$.  The frequencies of these orbits satisfy (\ref{eq:ResFrequenciesSec2}) of Lemma A. }   
\medskip\medskip\medskip

\noindent
{\em Proposition 3.2} \hspace{.05in} (Capture by $P_2$ implies weak capture)
\medskip

\noindent
{\em Let $P_0$ be captured with respect to $P_2$ at a distance $r$ from $P_2$ at a time $t_0$, where $E_2(\mathcal{Z}(t_0)) = E_{2R}(\zeta(t_0)) < 0$. Then, $P_0$ is weakly captured at $t_0$. That is, $P_0$ moves to weak escape at a time $t=t^*>t_0$, where $E_2(\mathcal{Z}(t^*)) =0, \hspace{.05in} E_2(\mathcal{Z}(t))  \gtrsim 0, t \gtrsim t^*$. (It is assumed Type IIa, Case B, $\Gamma$ points are excluded.)}
\medskip\medskip

\noindent
Proof of Proposition 3.2 - We distinguish several types of weak capture points.
\medskip\medskip

\noindent
Type I is where $P_0$ is on or near ${W}^u_k$ at $t= t_0$. In this case, $P_0$ is at a distance $r$ from $P_2$ where $\dot{r} = 0$ or $|\dot{r}| \gtrsim 0$,  where we have made use of the fact $W^u_k$ is open, so that $e_2 < 1$ and $E_2(t_0) < 0$. Thus, $P_0$ is captured at $t=t_0$.  For $t>t_0$, the proof follows by Proposition 3.1.
\medskip

\noindent  
Type II is where $P_0$ is not near ${W}^u_k$ since $|\dot{r}|$ is not near $0$ at $t=t_0$. There are two types. Type IIa is where $P_0$ starts at $t=t_0$ with $E_2$ and then to weak escape, with no cycling, by definition.    
If $P_0$ starts on a Type IIb point for $t=t_0$, then for $t > t_0$ there will be a time $\hat{t} > t_0$ where $ \dot{r} = 0$ or $|\dot{r}| \gtrsim 0$. In that case, $P_0$ is on or near $W^u_{k'}$ at $t=\hat{t}$, for some $k' \geq 1$. This yields a Type I point, that implies weak capture.    
\medskip

\noindent
In all these cases, $P_0$ moves to weak escape at a time $t=t^* > t_0$, where $E_{2R}(\zeta(t^*)) =0$, \hspace{.05in} $E_{2R}(\zeta(t))  \gtrsim 0, t \gtrsim t^*$.  This proves Proposition 3.2.
\medskip

\noindent 
As in \cite{BGT:2010}, we exclude Type IIa points as they are not generic. Points on $\Gamma$ are a set of measure $0$ and can be omitted.  Case B points are non-generic and excluded. 
\medskip\medskip\medskip

We now determine what kind of motion $P_0$ has about $P_2$ for times up to weak escape at $t^*$.   Consider the trajectory of $P_0$ as it undergoes counterclockwise cycling about $P_2$ after leaving points on or near $W^u_k$ on a line $\mathcal{L}$ in both Types I, IIb. (similar results are obtained for clockwise cycling) As $P_0$ performs $k-1$ cycles, it either has weak escape prior to completing the $k$th cycle, where $E_2=0$, and then when it intersects $\mathcal{L}$, $E_2 >0$, we call Case B, or it moves to cycle $P_1$ after the $(k-1)$st cycle where it was shown in Proposition 3.1 that $P_0$ achieves weak escape, we refer to as Case A. 
\medskip\medskip

\noindent
{\em Proposition 3.3}  ($P_0$ escapes from $P_2$ through the $N_1,N_2$ regions)  
\medskip

\noindent
{\em Assume $P_0 \in W^u_k$ at $t=0$, $C \lesssim C_2$, assuming Case A, and excluding Case B. Then after $(k-1)$-cycles about $P_2$, $P_0$ moves away from $P_2$, passing through the interior region of $M^s_2(\gamma_2)$ into $N_2$, between $H_2$ and $H_1$, through the interior of $M^u_1(\gamma_2)$, into $H_1$ where it starts to cycle $P_1$. When $P_0$ is within $N_2$, $E_2 \lesssim 0$.   If $C \lesssim C_1$, then after $(k-1)$-cycles about $P_2$, $P_0$ moves away from $P_2$ , passing through the interior of $M_2^s(\gamma_2)$ into $N_2$ between $H_2$ and $H_1$, and out into $H_1$ as before, or $P_0$ passes through the interior of $M_2^s(\gamma_1)$ into  $N_1$ between $H_2$ and $H_3$, and out into $H_3$ through the interior of $M_3^u(\gamma_1)$. When $P_0$ is within $N_1$, $E_2 \lesssim 0$.} 
\medskip\medskip

\noindent
Proof of Proposition 3.3 
\medskip

\noindent
Case A is considered with $C \lesssim C_2$. $P_0$ starts on $\mathcal{L}$ at $t = t_0$ with $E_2 < 0$, $\dot{r} =0$. It cycles about $P_2$, completes the $(k-1)$st cycle, then moves to $H_1$ where it starts to cycle about $P_1$, where $\dot{\theta}_1 >0$, for $0 \leq \theta_1 \leq 2\pi$ (see \cite{Belbruno:2004}). 
By the separatrix property $P_0$ is within the interior region contained by $M^s_2(\theta_2)$ on $\mathcal{L}$ at $t=t_0$ and it must pass from $P_2$, through $N_2$, where it is a transit orbit \cite{BGT:2010}.  When $P_0$ passes through $N_2$ it must pass inside the region bounded by $M_2^s(\gamma_2)$, and emerge from $N_2$ inside the region bounded by $M^u_1(\gamma_2)$ at $S^2_R$, where it will begin to cycle about $P_1$. For $\mu$ sufficiently small, the width of $N_2$ is near $0$, and geometrically this implies the velocity of $P_0$ with respect to $P_2$ is near $0$ since it passes close to $L_2$ in phase space. 
\medskip

\noindent
When $P_0 \in S^2_R$ in $H_1$, the distance from $P_0$ to $P_2$ can be estimated. The value of $C \lesssim C_2$, and $C_2  = 3 +  (\mu/3)^{2/3} + \mathcal{O}(\mu/3)$.  $P_2$ is near $L_2$. 
It directly follows  that $r \equiv r_2 = (\mu/3)^{1/3} + \mathcal{O}((\mu/3)^{2/3}))$. (This implies, $r_1 = 1 - |\mathcal{O}((\mu/3)^{1/3}))| \lesssim 1$ since $P_0$ is slightly to the right of $L_2$ at $S^2_R$.) 
\medskip

\noindent
The estimate of $r_2$ implies  that  for $\mu$ sufficiently small $r_2 \approx \mu/3$. Also, at $L_2$,  $\dot{x} =0$. Thus,  Equ. \ref{eq:KepEngRotate}  implies 
\begin{equation}
E_{2R} = (-3^{1/3}+(1/2)3^{-2/3})\mu^{-2/3} +\mathcal{O}(\mu^{b})  \lesssim 0,
\label{eq:E2Est}
\end{equation}
$b > 2/3$. 
\medskip

\noindent
Case A is considered with $C \lesssim C_1$. (As $r$ increases along $\mathcal{L}$, keeping a constant eccentricity, $C$ will decrease and move slightly below the other value, $C_1$ for $L_1$, $180$ degrees away from $L_2$ on the anti-$P_1$ side of $P_2$, $C \lesssim C_1$.) As $t$ increases from $t_0$, by the separatrix property, $P_0$ has two possibilities: (i) $P_0$ can pass through the region bounded by $M^s_2(\gamma_1)$, through $N_1$ and exiting within the region bounded by $M^u_3(\gamma_1)$ into $H_3$ intersecting $S^2_L = \partial N_1$ at a time $t_1$. It can then start to cycle about $P_1$ in $H_3$ for $t > t_1$, where $\dot{theta}_1 > 0$.  This implies unstable motion occurs, where after $(k-1)$-cycles about $P_2$, $P_0$ starts to cycle about $P_1$ in the $H_3$ region. It is similarly verified that (\ref{eq:E2Est}) is satisfied on $S^2_L$ at $t = t_1$.  This is different from when $P_0$ cycles about $P_1$ after $(k-1)$-cycles as it emerges from $N_2$ into the $H_1$ region.   However, in both cases, as seen, $E_2 \lesssim 0$ in the neck region bounding spheres. (ii) $P_0$ passes through $N_2$ into $H_1$. This yields the same results as in Case A.  
(It is verified that $C \lesssim C_1$ is sufficient to yield the same estimates in this proof for $E_{2R}$ as obtained for $C \lesssim C_2$.) 
\medskip\medskip

In summary, given $P_0 \in W^u_k$ at time $t = t_0$ there exists a time $t_1 > t_0$ where $E_2 \lesssim 0$ which occurs at $S^2_R \subset H_1$ for $C \lesssim C_2$, and for $C \lesssim C_1$, $E_2 \lesssim 0$ on $S^2_L \subset H_3$, or on $S^2_R \subset H_1$. Thus, in both cases, approximate parabolic escape occurs.
\medskip\medskip

\noindent
In the next step, we see what happens as $P_0$ starts to move about $P_1$ after leaving $S^2_R = \partial N_2$ in $H_1$, or $S^2_L = \partial N_1$ in $H_3$, through $M^u_1(\gamma_2)$, or $M^u_3(\gamma_1)$, respectively.  
\medskip\medskip

\noindent
{\em Proposition 3.4 }  ($P_0$ leaves $S^2_R$ ($S^2_L$), moves in approximate resonance orbit about $P_1$, returns to $S^2_R$ ($S^2_L$) and then to weak capture by $P_2$)
\medskip 

\noindent
{\em Assume $P_0 \in S^2_R \subset H_1$ at $t=t_1$. $P_0$ moves from $S^2_R$ for $t > t_1$ into an approximate resonance orbit about $P_1$. After $j$ cycles, $j \geq 1$, $P_0$ returns to $S^2_R$ where $E_2 \lesssim 0$. It then moves through $N_2$ to weak capture by $P_2$. 
\medskip

\noindent
(Similarly, assuming $P_0 \in S^2_L \subset H_3$ at $t=t_1$, $P_0$ moves from $S^2_L$ for $t > t_1$ into an approximate resonance orbit about $P_1$ in the outer Hills region $H_3$. After $j$ cycles, $j \geq 1$, about $P_1$, $P_0$ returns  $S^2_L$ where it then moves through $N_1$ to weak capture by $P_2$.) }
\medskip\medskip
   
\noindent
Proof of Proposition 3.4 - The case of $P_0 \in S^2_R \subset H_1$ is considered first, 
where $t= t_1$, $C \lesssim C_2$.  \cite{BGT:2013} is referenced since it determines the set $\mathcal{W}_k$ about 
the larger primary $P_1$ analytically. 
\medskip

\noindent
When $P_0 \in S^2_R$ for $t = t_1$, this implies it lies in the three-dimensional region bounded by $M_1^u$. Moreover, for $t > t_1$, due to the separatrix property, $P_0$ stays within this region inside $M_1^u$ for all time moving forward \cite{BGT:2013}. This manifold stays within a bounded region, $\mathfrak{M}_1$, bounded by the following: $S^2_R$, the boundary of $H_1$ (a zero velocity curve), and a two-dimensional McGehee torus, $T_M$, about $P_1$ \cite{BGT:2013}, \cite{LMS:1985}. \footnote{$T_M$ exists due to the fact that KAM tori on $\{ J=C \}$ cannot exist too close to $P_2$}.  The width of $\mathfrak{M}_1$ is $\mathcal{O}(\mu^{1/3})$. 
\medskip

\noindent
There are two cases. The first is where $M_1^u$ is a homoclinic two-dimensional tube which transitions from $M^u_1$ to $M_1^s$ which goes to $S^2_R$. This implies 
$P_0$ returns to $S^2_R$ at a later time. Now, if $M^u_1$ intersects $M^s_1$ transversally, then these manifolds intersect in a complex manner, where the image of $M_1^u$ on two-dimensional sections, $S_{\theta_1^0}$, are not circles, but parts of circles after several cycles of $P_0$ about $P_1$,  However, the separatrix property is still preserved, and $P_0$ still returns to $S^2_R$ \cite{BGT:2013}.

Let's assume it returns to $S^2_R$ after a time $T$, ($t_2 = t_1 +T$). $P_0$ is a transit orbit and must pass through $N_2$ for $t>t_2$ into $H_2$ through the interior region bounded by $M^u_2$, where it is again weakly captured by $P_2$. This follows since when $P_0$ passes through $N_2$ into $H_2$, within the interior region bounded by $M^u_2$, it will intersect $S^2_L \subset N_2$ in $H_2$. The estimate obtained in (\ref{eq:E2Est}) is also obtained at $S^2_L$.  This implies $P_0$ is captured by $P_2$ at $S^2_L \subset N_2$ at a time $t_2 + \delta$, $\delta > 0$. Under the previous assumptions on capture points in Theorem A, $P_0$  is weakly captured and weakly escapes $P_2$. 
\medskip

\begin{figure}
\centering
	\includegraphics[width=0.95\textwidth, clip, keepaspectratio]{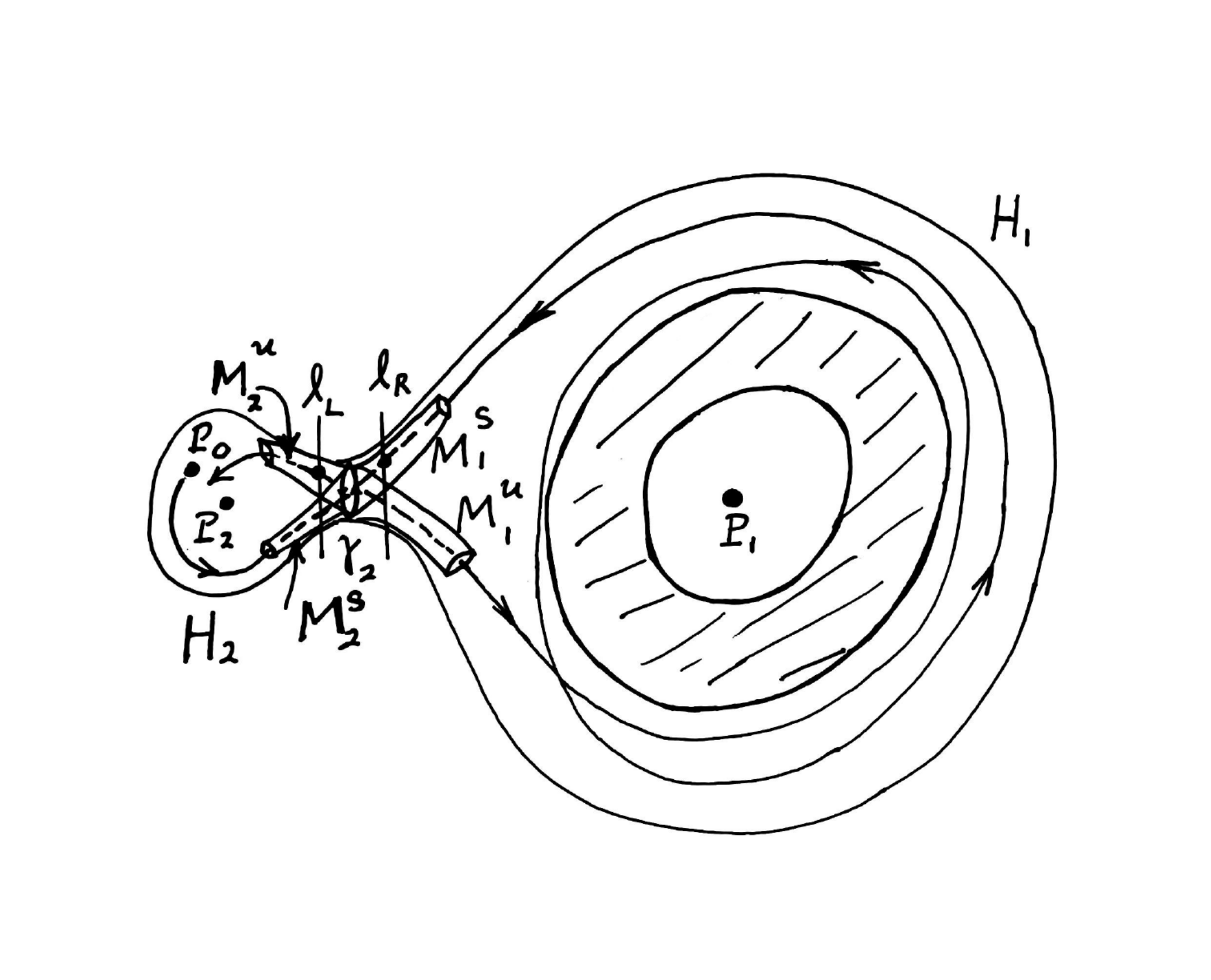}
\caption{The trajectory of $P_0$ is illustrated from leaving weak capture near $P_2$ at time $t_0$, passing into the neck $N_2$ from the interior three-dimensional region bounded by $M^s_2(\gamma_2)$, to the line $l_R$ at $t=t-1$, or equivalently the bounding sphere $ S^2_R = J^{-1}(l_R)$. $P_0$ then cycles about $P_1$ in $H_1$ within the region $\mathfrak{M}_1$ and eventually returns to $S^2_R$ at $t=t_2$ It then goes back into $H_2$ through $N_2$ and to weak capture relative to $P_2$ at $S^2_L$. This figure is a sketch. }
\label{fig:Fig3}
\end{figure}

\medskip
The motion of $P_0$ as it leaves weak capture near $P_2$, passing into the $H_1$ region and moving back to the $H_2$ to weak capture is illustrated in Figure \ref{fig:Fig3}.

\medskip

\noindent
 It is noted that there exists a time 
$\tilde{t}_3 < t_2$ where $E_2 > 0$, which follows from the proof of Proposition 3.1. Thus, $P_0$ is weakly captured in backwards time at $t=t_2 + \delta$. 
\medskip\medskip

\noindent
 A similar argument holds for $C \lesssim C_1$.  Within $H_2$ there are openings at $N_1$ to the left of $P_2$ and $N_2$ to the right. $P_0$ can now move into $H_3$ through $N_1$, in addition to moving into $H_1$ through $N_2$, from weak capture points on $W^u_k \cap \mathcal{L}$ in $H_2$ after $j$ cycles. If $P_0$ moves into $H_1$, it does so from the region bounded by $M_1^u(\gamma_2)$ and the same argument follows from the case $C \lesssim C_2$.  
\medskip

\noindent
If $P_0$ moves about $P_1$ in $H_3$, it moves in a bounded region $\mathfrak{M}_3$. This region is bounded by 
$S^2_L = \partial N_1$ in $H_3$, a McGehee torus $T_M$ about $P_1$ in $H_3$ and the boundary of $H_3$. 
$P_0$ moves inside the region enclosed by $M^u_3(\gamma_1)$ and stays within it as this manifold either transitions into 
$M^s_3(\gamma_1)$ as a homoclinic tube or if the manifolds have transverse intersection. The separatrix property is satisfied , and $P_0$ will cycle about $P_1$ in $H_3$ $j$ times until transits into $H_2$ and intersects $S^2_R = \partial N_1 $ in $H_2$, where (\ref{eq:E2Est}) is satisfied which implies $P_0$ is captured by $P_2$ at a time $t_2  + \delta$. Under the assumptions of Theorem A, $P_0$ is weakly captured at $P_2$.   
\medskip

\noindent
It is noted $P_0$ is weakly captured in backwards time, since there is a time $t_3 < t_2$ where  $E_2 > 0$, when $P_0$ was moving in $H_3$, using the same argument as in the proof of Proposition 3.1.
\medskip

\noindent
The final part of the proof of Proposition 3.4 is to prove that $P_0$ moves in resonance orbits about $P_1$.  
\medskip\medskip

\noindent
We consider $C \lesssim C_2$, where $P_0$ is moving in $H_1$.
$P_0$ is on an approximate resonance orbit in $H_1$ about $P_1$ for $t_1 \leq t \leq T$. This is proven as follows: $P_0$ moves in $\mathfrak{M}_1$. The orbit for $P_0$ will not deviate too much for $\mu$ sufficiently small, by the amount $\mathcal{O}(\mu^{1/3})$ \cite{LMS:1985}.  It is an approximate elliptic Keplarian orbit about $P_1$, since its energy $E_1<0$(proven in the following text, see Proposition A). It has a uniform approximate Keplerian period, $T_1$, for $\mu$ sufficiently small, with approximate frequency $\omega_1 = T_1^{-1}$. Once $P_0$ moves away from $S^2_R$ for $t > t_1$, and returns to $S^2_R$ for $t = t_2 = T + t_1$. When $P_0$ returns to $S^2_R$, it returns to near $P_2$ to approximately the distance $\mathcal{O}((\mu/3)^{2/3})$,  as follows from the proof of Proposition 3.3.
\medskip

\noindent
Since $P_0$ returns to $S^2_R$, near to $P_2$, $T$ must approximately be an integer multiple, $n$, of the period, $T_2$, of $P_2$ about $P_1$. That is,  $T \approx nT_2$. Also, since $P_0$ returns to near where it started, $T \approx mT_1$. Thus, $mT_1 \approx nT_2$.  Equivalently, $n\omega_1 \approx m\omega$. Thus, $P_0$ moves in an approximate $n:m$ resonance with $P_2$. It is noted that the approximate elliptic orbits of $P_0$ have an apoapsis distance from $P_1$ that is approximately the distance of $S^2_R$ to $P_1$. 
\medskip

\noindent
This can also be visualized in inertial coordinates, $(Y_1, Y_2)$ centered at $P_1$.  When $P_0$ has started its motion on a near ellipse, for $t > t_1$, it has just left weak capture from near $P_2$ at the location, $Y^*$. $P_0$ then cycles about $P_1$ and keeps returning to near $Y^*$ each approximate period $T_1$. When it arrives near $Y^*$, $P_2$ needs to be nearby as when $P_0$ started its motion. Otherwise, $P_0$ won't become weakly captured by $P_2$ and leave the ellipse to move to the $H_2$ to weak capture by $P_2$. In that case it will continue cycling about $P_1$.  If it does return to near $Y^*$ and $P_2$ has also returned near to where it started also near $Y^*$, then this means $P_2$ has gone around $P_1$ approximately $n$ times and $P_0$ has gone around $P_1$ approximately $m$ times.  
\medskip

\noindent
In the case where $P_0$ moves in the $H_3$ region after leaving $S^2_L$ for $t > t_1$, one also obtains a resonance orbit by an analogous argument. In this case, $P_0$ has a periapsis near $S^2_L = \partial N_1$ with respect to $P_1$, where  $S^2_L = \partial N_1$ is near $P_2$ at a distance of approximately, $\mathcal{O}((\mu/3)^{1/3})$.  These resonance orbits in $H_3$ are much larger than the resonance orbits in $H_1$ since they move about both $P_1, P_2$.
\medskip\medskip\medskip

\noindent
{\em Proposition A} \hspace{.05in} {\em $E_1 < 0$ when $P_0$ moves in $H_1$ about $P_1$ in a resonance orbit.}
\medskip

\noindent
Proof of Propisition A \hspace{.05in} When $P_0$ moves for $t > t_1$ it moves in an approximate two-body manner for finite time spans, where the osculating eccentricty $e_1$ and semi-major axis $a_1$ vary only for a small amount amount by $\mathcal{O}(\mu^{1/3})$ since $P_0$ moves within $\mathfrak{M}_1$. The energy $E_1$ is estimated(in an inertial frame). Since at $t = t_1$, $V_1(t_1) \approx 1-|z(t_1)|$. We can estimate $|z(t_1)|$ as roughly the distance of $L_2$ to $P_2$. This implies, $|z(t_1)| \approx \alpha^{1/3} + \mathcal{O}(\alpha^{2/3})$, $\alpha = \mu/3$, and $r_1 \approx 1 - \alpha^{1/3} + \mathcal{O}(\alpha^{2/3})$.  Thus,                                                   
\begin{equation}
E_1 \approx  (1/2)(1 - \alpha^{1/3} + \mathcal{O}(\alpha^{2/3}) )^2 - 
(1 - \mu) (1 - \alpha^{1/3} + \mathcal{O}(\alpha^2/3))^{-1} . 
\label{eq:Est}
\end{equation}
\medskip
Thus $E_1 \approx -(1/2) + \mathcal{O}(\alpha^{1/3})$. This implies that $E_1 < 0$ for $\mu $ sufficiently small. $P_0$ will then be moving on an approximate ellipse about $P_1$ of an eccentricity, $ e_1 < 1$.  The apoapsis of this ellipse will approximately be $ r_a \approx r_1(t_1) \approx 1 - (\mu/3)^{1/3} $. 
The semi-major axis of the ellipse at $t= t_1$ is approximately, $a_1 \approx -(1-\mu)/(2E_1) \approx (1 -\mu)+ \mathcal{O}((\mu/3)^{1/3})$. $e_1 \approx 1 - (r_a/a_1) < 1$. 
\medskip

\noindent When $P_0$ moves in resonance orbits in $H_3$ for $C \lesssim C_1$, similar estimates are made where $E_1 < 0$.  
\medskip

\noindent End of proof of Proposition A.
\medskip\medskip

\noindent
The resonance orbits of $P_0$ about $P_1$ move in an approximate two-body fashion where the perturbation due to $P_2$ is negligible for finite time spans. Thus, $\omega_1(t) \approx (m/n)\omega$ until it enters either neck to move to weak capture near $P_2$.
\medskip

\noindent 
We assume $C \lesssim C_2$ and examine what happens to the motion of $P_0$ near $Y^*$ when in $n:m$ resonance with $P_2$. $P_0$ is at a minimal distance to $P_2$ when near $Y^*$. In the rotating system, it is near $S^2_R$, and lies in the three-dimensional region contained within  $M_1^s$, since it has been moving about $P_1$ within this region by the seperatrix property. At this minimal distance, $M_1^s$ is close enough to $\gamma_2$ so that it can connect with it, and $P_0$ can move as a transit orbit and move through $N_2$ and exit into $H_1$ through the three-dimensional interior region bounded by $M_1^u$.  It is then captured by $P_2$, with $E_2 \lesssim 0$ at $S^2_L$.  An analogous arument holds when $C \lesssim C_1$.
\medskip

\noindent
Assuming the generic assumptions are satisfied for Theorem A, $P_0$ will weakly escape $P_2$ and again move into $H_1$, or $H_3$, obtaining  resonance orbits, satisfying, $\omega_1 \approx (m'/n')\omega$, for integers $n'\geq 1, m' \geq 1$. The set of all such resonance orbits forms the family $\mathfrak{F}$.  This concludes the proof of Theorem A.
\medskip\medskip\medskip\medskip\medskip\medskip

\begin{figure}
\centering
	\includegraphics[width=0.70\textwidth, clip, keepaspectratio]{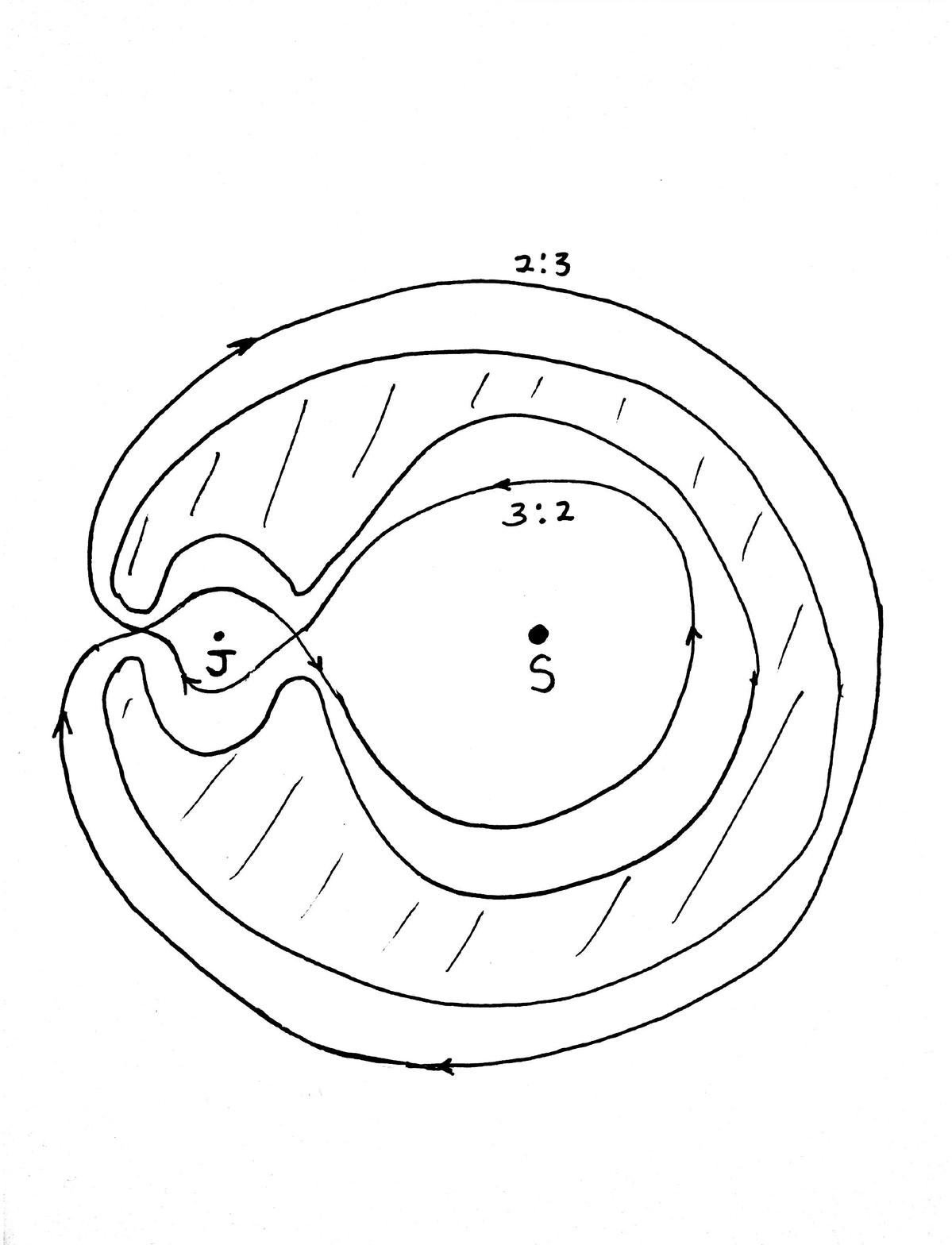}
\caption{A sketch showing the trajectory of the comet Oterma about the Sun(S) as it changes resonance types by weak capture near Jupiter(J) from $2:3$ to $3:2$, relative to the regions, $H_k$, $k=1,2,3$.   (Sketch and not to scale)  (see \cite{BelbrunoMarsden:1997}, \cite{KLMR:2001} for accurate Oterma  plot)}
\label{fig:Fig190}
\end{figure}

An example of the geometry of resonance transitions for an observed comet, Oterma, from a $2:3$ to a $3:2$, in 1936, and then back from a $3:2$ to a $2:3$, in 1962, ( \cite{BelbrunoMarsden:1997} , \cite{KLMR:2001} ) is illustrated in Figure \ref{fig:Fig190}. (see Section 3.1 at the end of this section.)

It is noted that the estimates of $E_1, E_2$ in the proof of Theorem 2 while $P_0$ moves in $H_i$, $i=1,2,3$, are observed 
in the motions of the resonating comets studied in \cite{BelbrunoMarsden:1997}. It can be seen in \cite{BelbrunoMarsden:1997} that when the comet Gehrels 3  was weakly captured by Jupiter($P_2$) from a $2:3$ resonance orbit into an approximate $3:2$ resonance orbit, $E_2 \lesssim 0$. Also, when the comet moved about the Sun($P_1$) in an approximate $2:3$ resonance orbit, $E_2 > 0$ and $E_1 < 0$. 
\medskip\medskip

For each resonance orbit obtained from the choice of the weak capture initial condition,  
(\ref{eq:ResFrequenciesSec2}) is satisfied, proving Lemma A.   
\medskip\medskip\medskip\medskip

\subsection{Examples of Resonance Orbits in $\mathfrak{F}$ and $\mathfrak{U}$.}
\medskip\medskip

Result A describes a dynamical mechanism of resonance orbits about $P_1$.   
The resonance motion described in this paper is observed both in nature and numerically.   

It was originally inspired by the fact that comets are observed to perform it. More exactly, there exists a special set of comets that move about the Sun that transition between approximate resonance orbits about the Sun due to weak capture at Jupiter.  This is studied in \cite{BelbrunoMarsden:1997},  \cite{KLMR:2001}, \cite{Ohtsuka:2008}.  Several comets are described in \cite{BelbrunoMarsden:1997}, \cite{Ohtsuka:2008},  that perform this motion. For example, the comet {\em Oterma} transitions between a $3 \mathbin{:} 2$-resonance with respect to the Sun, where $\omega_1 = (2/3)\omega$, $\omega$ is the frequency of Jupiter, to a $2 \mathbin{:} 3$-resonance. When passing between these resonances, the comet, $P_0$, is weakly captured by Jupiter. There are many others, listed in  \cite{BelbrunoMarsden:1997} (Table  1), and in \cite{Ohtsuka:2008}. These comets include Helin-Roman-Crockett ($3 \mathbin{:} 2 \rightarrow 3 \mathbin{:} 2$), Harrington-Abell ($5 \mathbin{:} 3 \rightarrow 8 \mathbin{:} 5$).  It is important to note that the modeling used to describe the resonance orbits of these comets is not exactly the model used in this paper. It models the true orbit of Jupiter about the Sun using the planetary ephemeris and the observed orbits of the comets for initial conditions which are not exactly planar. This model is very close to the planar restricted three-body problem.  The definition of approximate resonance orbits in this paper for the restricted three-body problem is well suited to the resonances comets perform.

The existence of orbits performing resonance transitions as in $\mathfrak{F}$ can also be found in the planar circular restricted three-body problem used in this paper. A special case where the resonance orbit precisely returns to its initial condition after preforming a transition was shown to exist in \cite{KLMR:2000}. This yields an exact periodic orbit that repeats the same transition over and over.
Other simulations of approximate resonance orbits as in $\mathfrak{F}$ for the planar circular restricted three-body problem and models very close to that model are done in \cite{Belbruno:1997}, \cite{BTG:2008}.  

An interesting example of orbits that occur in nature can be obtained for $\mathfrak{U}$ given in Result C. These are the subset of resonance transition orbits that have  frequencies, $\omega_1(m/n) \approx (m/n)\omega$, $m=8, n=\tilde{n}^3$, $\tilde{n} = 1,2, \ldots$. That is, the orbits have $\tilde{n}^3 \mathbin{:} 8$ resonances.   A special case of these resonances is an $8 \mathbin{:} 8$ resonance for $\tilde{n}=2$. On the other hand, a  $1 \mathbin{:} 1$ resonance orbit is a special case of an $8 \mathbin{:} 8$ resonance orbit. An example of this is for the Trojan asteroids, where $P_1$ is the Sun, $P_2$ is Jupiter, and $P_0$ is a Trojan asteroid. Many other examples can be found by asteroids located near the equilateral Lagrange points with respect to a body $P_2$, orbiting $P_1$.
\medskip\medskip\medskip\medskip

\section{Modeling Resonance Motions with the Modifed Schr\"odinger Equation}
\label{sec:SchroEqu} 
\medskip

\noindent
In this section some of the results are expanded upon in Section \ref{sec:Results}. 
\medskip\medskip

The family $\mathfrak{F}$  of resonance periodic orbits, $\Phi_{m/n}$, are modeled in the plane by the restricted three-body problem. The planar modeling is justified in Section \ref{sec:WeakCapture}.  To try and model $\mathfrak{F}$ with quantum mechanical ideas, we therefore use planar modeling. Thus, we consider the planar, time independent, modified Schr\"odinger equation given in the Introduction by (\ref{eq:Schrodinger}), obtained from the classical  Schr\"odinger equation by replacing $\hslash$ by $\sigma$, and the potential is given by the three-body potential $\bar{V}$ derived from the planar restricted three-body problem. This partial differential equation is time independent. 

The motivation of replacing $\hslash$ by $\sigma$ is given by Analogy A in Section \ref{sec:Results}, where $\sigma$ given by (\ref{eq:Sigma}). The potential $\bar{V}$ is given by (\ref{eq:VBar}), obtained from $\hat{V} =  V_1 + V_2$. It is recalled that $V_1$ is the potential due to $P_1$ and $V_2$ is the potential due to $P_2$, in an inertial $P_1$-centered coordinate system, $(Y_1, Y_2)$. It is also recalled that $P_2$ moves about $P_1$ on the circular orbit, $\gamma(t)$, $m_0 \gtrsim 0$, and as described in Section \ref{sec:Results} for Results A, it is necessary that $m_0, m_2$ are sufficiently small for the approximations in Assumptions 1.

As described in Section \ref{sec:Results} the modified Schr\"odinger equation is given by (\ref{eq:SchrodingerNew}), where $\bar{V}$ is the average of $\hat{V}$, obtained by averaging $V_2$ over a cycle of $P_2$ about $P_1$ on $\gamma$, given by (\ref{eq:AveragedPotentialSec2}).  For reference, we recall (\ref{eq:SchrodingerNew}),
\begin{equation}
-\frac{\sigma^2}{2\nu}\nabla^2 \Psi  + \bar{V} \Psi = E \Psi,
\label{eq:SchrodingerNewSec4}
\end{equation}
In the macroscopic scale for the masses, and relative distances, two dimensions is required, where, in the inertial $P_1$-centered coordinates, $(Y_1, Y_2)$  
$\nabla^2 \equiv \frac{\partial^2}{\partial Y_1^2} + \frac{\partial^2}{\partial Y_2^2}$.
When solving (\ref{eq:SchrodingerNewSec4}), the three-dimensional problem is solved for generality. The three-dimensional problem is 
discussed when considering the quantum scale.
\medskip

\noindent
It is noted that the units of $\sigma$ are $m^2 kg s^{-4/3}$. This needs to match the units of $\hslash$ which are $m^2 kg s^{-1}$.($\hslash = 6.62607 \times 10^{-34} m^2 kg s^{-1} $)  Thus, we need to multiply $\sigma$ by $\rho = 1 s^{1/3}$, $\sigma^* = \rho\sigma $. $\sigma^*$ has the same units as $\hslash$. Keeping the same notation,  $\sigma^* \equiv \sigma$.  At the end of this section it is seen that when the masses approach the quantum scale, $\sigma$ also gets small, and $m_0, m_1$ can be adjusted so that $\sigma = \hslash$.
\medskip\medskip

\noindent
We recall (\ref{eq:VBar}), $\bar{V}= V_1 + \bar{V}_2$, where $V_1 = -Gm_0m_1/r$, $r = |Y|$ and $\bar{V}_2$ is the time average of $V_2 = -Gm_0m_2/r_2$,  $r_2 = |Y - \gamma(t)|$. It is shown in this section, $\bar{V}_2$ is given by (\ref{eq:AllPotentialCases}) which has a form similar to $V_1$. This enables solving (\ref{eq:SchrodingerNew}).  
\medskip

\noindent
The solution to (\ref{eq:SchrodingerNewSec4}) is done in two steps. In the first step, we solve (\ref{eq:SchrodingerNewSec4}) in the absence of gravitational perturbations due to $P_2$ for $m_2 =0$, where $\bar{V}_2 = 0$. Thus, in this case $\bar{V} = V_1$. It is then solved for $m_2 > 0$ where $\bar{V}_2$ is non-zero.
(Since the form of  $\bar{V}_2$ is shown to have a form analogous to $V_1$, we can modify the solution obtained for $V_1$ for the addition of $\bar{V}_2$. ) 
\medskip

\noindent
We explicitly solve (\ref{eq:SchrodingerNewSec4}), with $m_2 =0$, by separation of variables. This is for the two-body motion of $P_0$ about $P_1$. It is first solved for three-dimensions, then restricted to the planar case studied in this paper for the macroscopic scale. The three-dimensional solution is referred to when discussing solutions in the quantum scale. (\ref{eq:SchrodingerNewSec4}) is transformed to spherical coordinates, $r, \phi, \theta$.  It is assumed the solution is of the form,  
\begin{equation}  
\Psi =  R(r) Y(\phi, \theta).
\label{eq:Sep1}
\end{equation}
$r \geq 0$, and $\phi$ is the angle relative to the $Y_1$-axis, $0 \leq \phi \leq 2\pi$. $\theta$ is the angle relative to the $Y_3$-axis, $0\leq \theta \leq \pi$. 
$|\Psi|^2$ is the probability of finding $P_0$ at distance $r$ from $P_1$.  
\medskip

\noindent
The solution of (\ref{eq:SchrodingerNewSec4}) for $\bar{V}_2 =0$ follows the method described in \cite{Atkins:2005} for the case of an electron in the Hydrogen atom moving about the nucleus.  This is a standard approach used in solving the classical Schr\"odinger equation in quantum mechanics found in many references.  There are some minor modifications. The Coulomb potential is used in \cite{Atkins:2005} for $V_1$ and here we are using the gravitational potential between two particles $P_0, P_1$; however, they are of the same form, both proportional to $r^{-1}$, $r^2 =Y_1^2 + Y_2^2 + Y_3^2$. Instead of the proportionality term of $Gm_0m_1$, the Coulomb potential has the term, $Ze^2c_1/(4\pi\epsilon_0)$, where $Z$ is the atomic number, $Z=1$ for Hydrogen, $e$ is the charge of the electron and the charge of the nucleus, $P_0$ and an atomic nucleus, $P_1$, $\epsilon_0$ is the permittivity of vacuum, $\hslash$ is replaced by $\sigma$. The reduced mass $\nu = m_0m_1/(m_0+m_1)$ is defined for either the gravitational or Coulomb modeling.  When referring to \cite{Atkins:2005}, one replaces $e^2/(4 \pi \epsilon_0)$ by $Gm_0m_1$.
\medskip

\noindent 
When solving (\ref{eq:SchrodingerNewSec4}) by separation of variables, (\ref{eq:Sep1}) is substituted into 
(\ref{eq:SchrodingerNewSec4}), obtaining differential equations for $R(r)$ and $Y(\phi, \theta)$,
\begin{equation}
\frac{d^2 R}{dr^2} + 2r^{-1}\frac{dR}{dr} + [2\nu \sigma^{-2} (E + Gm_0m_1 r^{-1}) - \alpha r^{-2}] R= 0,
\label{eq:RDE}
\end{equation} 
\begin{equation}
(\sin \theta)^{-1}\frac{\partial}{\partial\theta}(\sin \theta \frac{\partial Y}{\partial \theta}) + 
(\sin \theta)^{-2} \frac{\partial^2 Y }{\partial \phi^2} + \alpha Y = 0  .
\label{eq:AngleDE}
\end{equation}
$\alpha$ is a separation constant.
\medskip

\noindent (\ref{eq:AngleDE}) is solved first, yielding spherical harmonics. Using separation of variables, solutions are obtained in the form, $Y(\phi, \theta) = \Phi(\phi)\Theta(\theta)$. This gives the differential equations,
\begin{equation}
\Theta^{-1} \sin \theta \frac{d}{d\theta}(\sin \theta \frac{d\Theta}{d\theta}) + \alpha \sin^2\theta -\beta =0,
\label{eq:ThetaDE}
\end{equation}
\begin{equation}
{{d^{2}\Phi}\over{d\phi^2}} + \beta \Phi = 0,
\label{eq:PhiDE}
\end{equation}
where $\beta$ is a separation constant \cite{Atkins:2005}. 
\medskip

\noindent (\ref{eq:PhiDE}) gives the solution, 
\begin{equation}
\Phi(\phi) \equiv \Phi_{m_l}(\phi) = (1/2\pi)^{1/2} \e^{im_l\phi}  , 
\label{eq:PhiSolution}
\end{equation}
$\beta = m_l^2, m_l = 0, \pm 1, \pm 2, \ldots$,  $i^2 = -1$. $\Phi(\phi)$ varies in the $Y_1, Y_2$-plane. 
\medskip

\noindent The solution of  (\ref{eq:ThetaDE}) follows by setting $x = \cos \theta$, $G(x) \equiv \Theta(\cos x)$  transforming (\ref{eq:ThetaDE}) into an associated Legendre type differential equation,
\begin{equation}
(1 - x^2)\frac{d^2 G}{dx^2} - 2x \frac{dG}{dx} + (\alpha - m^2(1-x^2)^{-1})G = 0,
\label{eq:LegendreDE}
\end{equation}
$\alpha = l(l+1), l = |m_l|,|m_l|+1,|m_l|+2, \ldots$ \cite{Atkins:2005}. $l$ varies between $\pm |m_l|$.   The solutions of (\ref{eq:LegendreDE}) are given by associated Legendre polynomials $P_{l,|m_l|}(x)$. (see \cite{Abramowitz:1974} for tables of these polynomials) The solutions of (\ref{eq:ThetaDE}) are given by (\cite{Atkins:2005}, page 527),  
\begin{equation}
\Theta_{l,m_l}(\theta) = \Big{[}\Big{(}\frac{2l+1}{2}\Big{)}\frac{1-|m_l|!}{1+|m_l|!}\Big{]}^{1/2}P_{l,|m_l|}(\cos \theta).
\label{eq:ThetaSolution}
\end{equation}
It is remarked that in the two-dimensional problem, with coordinates $(Y_1, Y_2)$, $\theta = \pi/2$. In this case, there is no variation with respect to $\theta$ and $\Theta$ is only defined at $\theta = \pi/2$. 
\medskip

\noindent
The solution $Y(\phi, \theta)$ is given by the spherical harmonics $Y_{l,m_l} = \Phi_{m_l}(\phi) \Theta_{l,m_l}(\theta)$.
\medskip

\noindent
To solve (\ref{eq:RDE}), set $u = Rr$. (\ref{eq:RDE}) becomes,
\begin{equation}
\frac{d^2 u}{dr^2} +  (\tilde{a}  r^{-1} - b r^{-2}) u = \lambda^2 u, 
\label{eq:u-equation}
\end{equation}
\medskip
where $\lambda^2 = 2\nu |E| \sigma^{-2}$, $\tilde{a} = 2\nu \sigma^{-2} Gm_0m_1$, $b = l(l+1)$.  It is verified that solving this differential equation yields the solution of (\ref{eq:RDE}), 
\medskip\medskip

\begin{equation}
R \equiv R_{\tilde{n},l}(r) = -\Big{[}\frac{2}{\tilde{n}}\Big{(}\frac{(\tilde{n}-l-1)!}{2\tilde{n}[(\tilde{n}+l)!]^3}\Big{)} \Big{]} \rho^l L^{1l+1}_{\tilde{n}+l}(\rho)\e^{-\rho/2},
\label{eq:R-detailed}
\end{equation}
where, $\tilde{n} = 1,2 \ldots$, $l = 1, \ldots , \tilde{n}-1$, $\rho=(2/(\tilde{n}a) ) r$, $a=(\nu Gm_0m_1)^{-1}\sigma^2$, and $L^i_j$ are the associated Laguerre polynomials (\cite{Atkins:2005}, page 8, Table 3.2), and 
\begin{equation}
E \equiv \hat{E}_{\tilde{n}} = - { {2\nu \pi^2 (Gm_0m_1)^2}\over{\sigma^2 \tilde{n}^2} }  .
\label{eq:E_n}
\end{equation}
\medskip
\noindent
$E$ can be written as,
\begin{equation}
\hat{E}_{\tilde{n}} = -\frac{4 \sigma}{\tilde{n}^{2}}.
\label{eq:EReduced}
\end{equation}
This follows by the identity, 
\begin{equation}
2\nu \pi^2 \tilde{\rho}^2 \sigma^{-3} = 4,
\label{eq:Identity}
\end{equation}
where $\tilde{\rho} = Gm_0m_1$.
\medskip\medskip

\noindent The solution for $E$ yields quantized values of the energy, which quantizes the gravitational field between $P_0$, $P_1$.  
\medskip

\noindent
The general solution to (\ref{eq:SchrodingerNewSec4}) is given by multiplying (\ref{eq:R-detailed}) with $Y_{l,m_l}$,
\begin{equation}
\Psi(r, \phi, \theta) \equiv \Psi_{\tilde{n},m_l,l} =  R_{\tilde{n},l}(r) \Phi_{m_l}(\phi) \Theta_{l,m_l}(\theta).
\label{eq:SolnGen}
\end{equation}
\medskip

\noindent
$\tilde{n}, l, m_l$ are the quantum numbers associated with $\Psi$. $\tilde{n}$ is called the principle quantum number and is independent of $l, m_l$.  It specifies the energy value and limits the value of $l$. The quantum numbers $l,m_l$ occur by consideration of the spherical harmonics. 
\medskip\medskip

\medskip
We compute the  probability distribution function, $F$, of locating $P_0$ relative to $P_0$ at a given point $(r, \phi, \theta)$. Since we are currently considering the masses and the relative distances to be macroscopic, this probability is used to measure the location of a macroscopic particle, and not as a wave as is done in the quantum scale, that is considered later in this section.  By definition, $F(r, \phi, \theta) = |\Psi_{\tilde{n},m_l,l}(r, \phi, \theta)|^2$, depending on the quantum numbers.  
\medskip

\noindent
It is more convenient to compute the probability at a given radial distance $r$, independent of $\phi, \theta$.  It is labeled $P(r)$.
\medskip

 \noindent
 It is verified that  $P(r) = R^2r^2$. This is valid in the two-dimensional case as well for $(Y_1, Y_2)$.
\medskip

\noindent
As an example, calculate $P(r)$ at the lowest energy value corresponding to $\tilde{n}=1$, $l=0$.
\begin{equation}
P(r) = 4 a^{-3}r^2 \e^{-2r/a}.   
\end{equation}   
(see \cite{Atkins:2005}, Table 3.2, where $a$ is given in the Hydrogen atom case).  It is verified that $P(r)$ has a maximum at $r = a$, with $P(0) =0$ and where $P(r) \rightarrow 0$ as $r \rightarrow \infty$.  It yields a curve $\{ (r,P(r))| r \in [0,\infty]\}$ analogous to the Hydrogen atom case (see \cite{Atkins:2005}, Figure 3.20). The numerical values of $P(r)$ will differ from the Hydrogen atom case, since in the gravitational case $a = (\nu Gm_0m_1)^{-1}\sigma^2$. 
\medskip

\noindent
The maximum of the distribution function at $r=a$ says that $P_0$, as a macroscopic body, has the highest probability of being located at this distance. In the case of the Hydrogen atom, where $P_0$ has wave-particle duality,  this distance corresponds to the Bohr radius, which is the most probable location to find an electron in general, referred to as the $1s$-orbital ($s$ denotes $l=0$).
\medskip

\noindent
In the same way, $P(r)$ can be computed for $\tilde{n} = 1,2 \ldots, \hspace{.05in} l = 0,1,2, \ldots, \tilde{n}-1$, which determines most probable radial locations for $P_0$ to be located.  
\medskip

\noindent If one considers computing $F(r, \phi, \theta) = |\Psi_{\tilde{n},m_l,l}(r, \phi, \theta)|^2$ over $\tilde{n}, l, m_l$, where $-|m_l| \leq l \leq |m_l|$, then one obtains precise regions about $P_1$ where $P_0$ is most probable to be located.  These are well known in the case of the Hydrogen atom \cite{Atkins:2005}. It is remarkable these are observed to occur. In the gravitational case considered in this paper, the regions will have a similar geometry, but with different scaling.
\medskip\medskip\medskip

The solution of (\ref{eq:SchrodingerNewSec4}) has been obtained in three dimensions for $m_2=0$. We solve this for $m_2> 0$, $m_2\ll m_1$, and then reduce to the planar case in order to compare to the planar restricted three-body problem in the macroscopic scale. 
\medskip

\noindent
When the gravitational perturbation due to $P_2$ is included, the previous results are obtained with a small perturbation. It is also seen that the frequencies $\omega_1(\tilde{n})$ correspond to the subset, $\mathfrak{U}$, of the resonant family $\mathfrak{F}$. 
\medskip\medskip\medskip

\noindent
{\em Three-Body Potential}
\medskip

  The previous analysis can be done for a more general three-body potential by taking into account the gravitational perturbation due to $P_2$. We do this by using an averaged potential, $\bar{V}_2$, obtained from the potential $V_2$, due to the gravitational interaction of $P_0,P_2$. This yields the three-body potential,  $\bar{V} = V_1 + \bar{V}_2$ that approximates $\hat{V} = V_1 + V_2$.  This is done as follows,
\medskip

\noindent	
We do this analysis in three-dimensional intertial coordinates, $Y = (Y_1, Y_2, Y_3)$, centered at $P_1$, $P_2$ moves about $P_1$ on a circular orbit of radius $\beta$, and angular frequency $\omega$,  in the $(Y_1, Y_2)$-plane,
$\gamma(t) = \beta(\cos \omega t, \sin \omega t, 0)$, $\beta$ is a constant. The potential for $P_0$ due to the perturbation of $P_2$ is given by
\begin{equation}
\hat{V}  =  V_1 + V_2 = \hspace{.08in} -\frac{Gm_0m_1}{r}  - \frac{Gm_0m_2}{r_2},
\end{equation}
where $r = |Y|, r_2 = |Y - \gamma(t)|$. We can write $r_2$ as
$$ r_2 = \sqrt{r^2 + \beta^2 - 2\beta(Y_1\cos\omega t + Y_2 \sin \omega t )},$$  
We consider $V_2$ and take the average of it over one cycle of $P_2$ , where $t \in [0, 2 \pi/\omega]$,
\begin{equation}
\bar{V}_2 = -\frac{Gm_0m_2\omega}{2\pi} \int_0^{\frac{2\pi}{\omega}} \frac{dt}{\sqrt{r^2 + \beta^2 -2\beta(Y_1 \cos \omega t + Y_2 \sin \omega t)}}.
\end{equation}
This averaged potential term is an approximation to $V_2$, representing the average value of $V_2$ felt by $P_0$ at a point $(Y_1, Y_2, Y_3)$ over the circular orbit of $P_2$ about $P_1$ in the $(Y_1,Y_2)$-plane. It is advantageous to use since it eliminates the time dependence in $V_2$, and as we'll show, can be written so that it approximately takes the form of $V_1$. This implies we can solve (\ref{eq:SchrodingerNewSec4}) as before, with minor modifications.  Approximating $V_2$ in this way yields $\bar{V}_2$.  
\medskip

\noindent
Expressing $Y_1, Y_2$ in  polar coordinates, $Y_1 = r \cos \theta, Y_2 = r \sin \theta$, and making a change of the independent variable, $t$,  $\phi = \omega t$, we obtain
\begin{equation}
\bar{V}_2 = -\frac{Gm_0m_2}{2\pi} \int_0^{2\pi} \frac{d\phi}{\sqrt{r^2 + \beta^2 -2\beta r\cos(\phi-\theta)}}.
\label{eq:AveragedPotential}
\end{equation}
\medskip\medskip

$\bar{V}_2$ is simplified by considering three cases, $r < \beta$, and $r > \beta$, $r = \beta$, and expanding $\bar{V}_2$ as a binomial series.

We prove,
\medskip\medskip

\noindent
{\em SUMMARY A}
\medskip

\noindent
The general three-body potential $\hat{V} = V_1 + V_2$, (\ref{eq:GeneralPotential}), for $P_0$ can be approximated by 
replacing $V_2$, due to the perturbation of $P_2$, with the averaged potential $\bar{V}_2$, (\ref{eq:AveragedPotential}).
$\bar{V}_2$ can be written as,  
\begin{equation}
\bar{V}_2 = 
  \left\{ \begin{array}{ll}
 -Gm_0m_2{r^{-1}} + \mathcal{O}(m_0m_2), &  r > \beta  \nonumber \\   
-Gm_0m_2{\beta^{-1}} + \mathcal{O}(m_0m_2), &  r < \beta \\  \label{eq:AllPotentialCases}
-Gm_0m_2\frac{1}{\sqrt{2}}r^{-1} + \mathcal{O}(m_0m_2), &  r = \beta. \nonumber
	\end{array}\right.
% \label{eq:AllPotentialCases}
   \end{equation}
The quantized energy, $\hat{E}_{\tilde{n}}$,  for the approximated three-body potential $\bar{V} = V_1 + \bar{V}_2$ is given by,
\begin{equation} 
\hat{E}_{\tilde{n}} = 
\left\{\begin{array}{ll}
 -{4 \sigma}\tilde{n}^{-2}(1 + \mu + \mu^2)+ \mathcal{O}(m_0m_2) , &  r > \beta \nonumber \\
-{4 \sigma}{\tilde{n}^{-2}}   - {Gm_0m_2}{\beta^{-1}} + \mathcal{O}(m_0m_2), &  r < \beta   \label{eq:AllEnergyCases}\\ 
 -{4 \sigma}{\tilde{n}^{-2}}(1 + \frac{1}{\sqrt{2}}\mu + \frac{1}{2}\mu^2)+ \mathcal{O}(m_0m_2), &  r = \beta  \nonumber
\end{array}\right.
%\label{eq:AllEnergyCases}
\end{equation}
which reduces to (\ref{eq:EReduced}) for $m_2 = 0$, and where $\mu = m_2/m_1$.   
\medskip\medskip
\medskip\medskip

 From the form of $\sigma$,  (\ref{eq:AllEnergyCases}) implies,   
\begin{equation}
\hat{E}_{\tilde{n}} =  -{4 \sigma}\tilde{n}^{-2} + \mathcal{O}(m_0m_2) .
\label{eq:SchroEnergy2}
\end{equation}
\medskip\medskip\medskip

\noindent
Similarly, 
\medskip\medskip\medskip\medskip

\noindent
{\em SUMMARY B}
\medskip

\begin{equation}
R_{\tilde{n},l}(r) = -\Big{[}\frac{2}{\tilde{n}}\Big{(}\frac{(\tilde{n}-l-1)!}{2\tilde{n}[(\tilde{n}+l)!]^3}\Big{)} \Big{]} \rho^l L^{1l+1}_{\tilde{n}+l}(\rho)\e^{-\rho/2} + \mathcal{O}(m_0m_2),
\label{eq:R-detailedwithpert}
\end{equation}
for $r \geq \beta, r < \beta$.  The probability distribution function is generalized to,
\begin{equation}
P(r) =  R^2(r)r^2   + \mathcal{O}(m_0m_2). 
\label{eq:RadialProbWithPert}
\end{equation}
\medskip\medskip\medskip
 																	
\noindent

\medskip
When adding the gravitational perturbation due to $P_2$ represented by $\bar{V}_2$, one obtains smooth dependence on this term in all
the calculations.
  This proves summary B.
\medskip\medskip\medskip

\noindent
{\em Equivalence of Solutions of the  Modified Schr\"odinger Equation with the Family $\mathfrak{F}$ of the Three-Body Problem} 

\medskip

The two-dimensional case is now considered to compare the solutions of the modified Schr\"odinger equation, (\ref{eq:SchrodingerNewSec4}), to the family of solutions $\mathfrak{F}$ of  the planar restricted three-body problem. 
\medskip

\noindent
We set $Y_3=0$, $\theta =\pi/2$. The quantized energy $\hat{E}_ {\tilde{n}}$, (\ref{eq:EReduced}), for (\ref{eq:SchrodingerNewSec4}) of the two-body motion of $P_0$ about $P_1$, with $m_2=0$, is not defined in the same way as the two-body Kepler energy $\tilde{E_1}$, (\ref{eq:Planck2}). $\hat{E}_ {\tilde{n}}$ is computed from the modified Schr\"odinger equation and $\tilde{E_1}$ is computed for the Kepler problem for general elliptic motion of $P_0$ about $P_1$. These are different expressions. However, they both represent the energy of $P_0$ for the two-body gravitational potential.   Also, $\hat{E}_ {\tilde{n}}$ is quantized and  $\tilde{E_1}$ is not quantized.     
\medskip

\noindent
A key observation of this paper is that when one solves for the Kepler frequency $\omega_1$ in (\ref{eq:Planck2}) as a function of $\tilde{E_1}$ and substitutes $\hat{E}_ {\tilde{n}}$ in place of  $\tilde{E_1}$, a simple equation is obtained for $\omega_1$,
\begin{equation}
\omega_1|_{\tilde{E_1} =  \hat{E}_ {\tilde{n}}}   \equiv  \omega_1(\tilde{n})  =  8 \tilde{n}^{-3}. 
\label{eq:FrequencyCalc}
\end{equation}
It is remarked that this equation is also valid in three-dimensions since (\ref{eq:Planck2}) is also valid for the three-dimensional Kepler two-body problem. 
\medskip

\noindent
This follows by noting that (\ref{eq:Planck2}) implies,
\begin{equation}
\omega_1 = [-\sigma^{-1}\tilde{E}_1]^{3/2} .
\label{eq:Frequency}
\end{equation}
This yields (\ref{eq:FrequencyCalc})
\medskip

\noindent Inclusion of gravitational perturbation of $P_2$ ($m_2 > 0$), implies more generally,
\begin{equation}
\omega_1|_{\tilde{E_1} =  \hat{E}_ {\tilde{n}}}   \equiv  \omega_1(\tilde{n})  =  8 \tilde{n}^{-3} + \mathcal{O}(m_0m_2). 
\label{eq:FrequencyCalcPert}
\end{equation}
\medskip

\noindent
(\ref{eq:FrequencyCalc}) does not depend on the masses or any other physical parameter. This says substituting the quantized energy from the modified Schr\"odinger equation, with $m_2 =0$, into the Kepler energy for the frequency, yields a frequency of motion for $P_0$ moving about $P_1$ in elliptical orbits that is same for all masses only depending on the wave number $\tilde{n}$.  This discretizes the Kepler frequencies.
\medskip

\noindent
The discretization of the Kepler frequencies restricts the elliptical two-body motion of $P_0$.  This result becomes relevant in the three-body problem for $\mathfrak{F}$ when the gravitational perturbation $P_2$ is included since it selects a the set of resonance orbits, $\mathfrak{U} \subset \mathfrak{F}$ (see Result C, Section \ref{sec:Results}).  
\medskip\medskip

We prove Result C from Section \ref{sec:Results}, which we state as
\medskip\medskip

\noindent
{\em  $\omega_1(\tilde{n})$, given by (\ref{eq:FrequencyCalcPert}), which are the frequencies in the restricted three-body problem corresponding to the modified Schr\"odinger equation energies, form a subset $\mathfrak{U} \subset \mathfrak{F}$ of resonance orbits where $m=8, n=\tilde{n}$.}
  \medskip\medskip
		
\noindent
{\em Proof of Result C}
\medskip

\noindent
This is proven by first noting that the circular restricted three-body problem can rescaled so that $m_1 = 1- \mu, m_2 = \mu, G =1, \beta =1$ where $\mu = m_2/(m_1+m_2)$ \cite{Belbruno:2004}. This scaling does not reduce the generality of the mass values nor $\beta$. This scaling implies $\omega =1$. Thus, (\ref{eq:ResFrequenciesSec2}) becomes,
\begin{equation}
\omega_1(m/n) =  (m/n) + \mathcal{O}(\delta).
\label{eq:ResFrequenciesSec2Reduced}
\end{equation}
A key observation is that this scaling does not effect the leading term ${8}/{\tilde{n}^{3}}$ of $\omega_1(\tilde{n})$ given by (\ref{eq:FrequencyCalcPert}). Thus, after the scaling, subtracting (\ref{eq:FrequencyCalcPert}) from (\ref{eq:ResFrequenciesSec2Reduced}) yields,
\begin{equation}
\omega_1(m/n) - \omega_1(\tilde{n}) = \frac{m}{n} - \frac{8}{\tilde{n}^3}  + \mathcal{O}(m_0m_2).
\label{eq:FrequenciesEquate}
\end{equation}
Thus, taking $m = 8, n = \tilde{n}^3$ and assuming $m_2$ is sufficiently small, implies,
\begin{equation}
\omega_1(m/n) \approx \omega_1(\tilde{n}),
\label{eq:approx}
\end{equation}
 This condition is preserved by rescaling to general $m_1, m_2, \beta$, yielding $\omega_1(m/n) \approx (8/\tilde{n}^3)\omega$ for $\mathfrak{U}$.  
\medskip\medskip\medskip

\noindent
{\em  From the Macro to Quantum Scale} 
\medskip\medskip

When $m_0, m_2, m_3$ are in the macroscopic scale  then as described in Section 
\ref{sec:Results}, in Result B and Result C, the family $\mathfrak{U} \subset \mathfrak{F}$ of near resonance orbits can be described by the solution (\ref{eq:wave}) of (\ref{eq:SchrodingerNewSec4}).  

For the masses in the quantum scale, the family $\mathfrak{U}$ are no longer valid. $\Psi$ given by (\ref{eq:wave}) is still valid but now as pure wave solutions.  This is summarized in Result D, Section \ref{sec:Results}. Thus, $\Psi$ is defined for both macroscopic and quantum scales. In the macroscopic scale, $\Psi$ is interpreted as a probability, whereas in the quantum scale, $\Psi$ is a pure wave solution.  The quantized energies $\hat{E}_{\tilde{n}}$ are still well defined. $P$ is still defined.
\medskip\medskip\medskip

This section is concluded with an analysis of $\sigma$, referred to in Section \ref{sec:Results}. 
\medskip

\noindent
{\em Proposition 4.1} \hspace{.05in}  $\sigma = \hslash$ is satisfied for a one-dimensional algebraic curve (\ref{eq:sigmaPlanck}) in $(m_0, m_1)$-space.
\medskip\medskip

\noindent
This is proven by noting 
$ \sigma = (1/2)(2 \pi G)^{2/3}m_0m_1(m_0 + m_1))^{-1/3} <  (1/2)(2 \pi G)^{2/3}m_0m_1m_0^{-1/3}$. Hence, $\sigma  < (1/2)(2 \pi G)^{2/3}m^{2/3}_0m_1$.
Thus, $\sigma \rightarrow 0$ as $m_0, m_1 \rightarrow 0$.
This implies there exists values of $m_0, m_1$ such that $\sigma = \hslash$. This is equivalent to the equation,
\begin{equation}
 m^3_0m^3_1 - \hslash a^{-3} (m_0 + m_1) = 0 , 
\label{eq:sigmaPlanck}
\end{equation}
$a = (1/2)(2 \pi G)^{2/3}$. (\ref{eq:sigmaPlanck}) yields a one-dimensional algebraic curve, $\Gamma$, in the coordinates $m_0, m_1$. This proves Proposition 4.1. 
\medskip \medskip

\noindent
The relevancy of possible wave motions for $(m_0, m_1)$ not on $\Gamma$, with $m_0, m_1, m_2$ in the quantum scale, is not considered in this paper. This is discussed in Section \ref{sec:Results}.
Different models are also discussed in Section \ref{sec:Results}.
\medskip\medskip
\medskip\medskip\bigskip

%\section{Conclusion}
%\label{sec:Conclusion} 
%\medskip

\ack

I would like acknowledge the support of Alexander von Humboldt Stiftung of the Federal Republic of Germany that made this research possible, and the support of the University of Augsburg for my visit from 2018-19. I would like to thank Urs Frauenfelder of the University of Augsburg  for many interesting discussions. Research by E.B. was partially supported by NSF grant DMS-1814543. Special thanks to Marian Gidea of Yeshiva University for helpful discussions and Figure 5. I would like to thank David Spergel of Princeton University. 
\medskip\medskip\medskip\medskip
\medskip\medskip\medskip\medskip

\noindent
{\bf  APPENDIX A} \hspace{.1in} {\em  Proof of SUMMARY A}
\medskip\medskip

 Summary A is proven as follows:   
\medskip

\medskip
The integrand, $I$, of $\bar{V}_2$ is given by
\begin{equation}
I =  \frac{1}{\sqrt{r^2 + \beta^2 -2\beta r\cos(\phi-\theta)}} = \frac{1}{ \sqrt{r^2 + \beta^2}} \frac{1}{\sqrt{1 - \frac{2\beta r}{r^2 + \beta^2}\cos(\phi-\theta) }}   .
\label{eq:ratio}
\end{equation}
\medskip\medskip\medskip

\noindent
Case 1:  $r > \beta$
\medskip

\noindent
This implies,
\begin{equation}
\frac{1}{\sqrt{r^2 + \beta^2}} =  \frac{1}{r\sqrt{1+ \frac{\beta^2}{r^2}}} = \frac{1}{r} (1 + \mathcal{O}(x)),
\label{eq:FirstRatio}
\end{equation}						
where	$x = \beta/r < 1$. This results from expanding the fraction containing the square root into a binomial series. Likewise, we can also expand the first term on the right in (\ref{eq:ratio}) in a binomial expansion since $|\cos(\phi-\theta)| \leq 1$  and
\begin{equation}
	\frac{2\beta r}{r^2 + \beta^2}   <  1,
\label{eq:otherterm}
\end{equation}
yielding 
\begin{equation}
\frac{1}{\sqrt{1 - \frac{2\beta r}{r^2 + \beta^2}\cos(\phi-\theta) }} =  1 + \mathcal{O}(y),
\end{equation}
where 
\begin{equation}
y = \frac{2\beta r}{r^2 + \beta^2}(\cos(\phi-\theta)),
\end{equation}
$|y| < 1$. Thus, (\ref{eq:ratio}) becomes,
\begin{equation}
I = \frac{1}{r} (1 + \mathcal{O}(w)),
\end{equation}
$|w| < 1$, $ w = \max\{ x, y\}$.  Thus, in this case,
\begin{equation}
\bar{V}_2 = -\frac{Gm_0m_2}{r}(1 + \mathcal{O}(w)) = -\frac{Gm_0m_2}{r} + \mathcal{O}(m_0m_2).
\label{eq:PotCase1}
\end{equation}  

\noindent This implies in the derivation of $\hat{E}_{\tilde{n}}$, we proceed as before and replace the numerator $Gm_0m_1$ of $V_1$ in (\ref{eq:RDE}) with $Gm_0(m_1 + m_2)$ and adding $\mathcal{O}(m_0m_2)$ to this term. This yields,
\begin{equation}
\hat{E}_{\tilde{n}} = - { {2\nu \pi^2 (Gm_0(m_1 + m_2))^2}\over{\sigma^2 \tilde{n}^2} } + \mathcal{O}(m_0m_2)
\label{eq:NewEnergyCase1}
\end{equation}
This can be written as, 
\begin{equation}
\hat{E}_{\tilde{n}} = -\frac{4 \sigma}{\tilde{n}^2}(1 + \mu + \mu^2)+ \mathcal{O}(m_0m_2),
\label{eq:NewEnergyCase1Simplified}
\end{equation}
where $\mu = m_2/m_1$, and where we have used (\ref{eq:Identity}).
\medskip\medskip

\noindent
Case 2: $r < \beta$
\medskip

    This case is done in a similar way as in Case 1. Instead of factoring out $r^{-1}$ from (\ref{eq:FirstRatio}), we factor out $\beta^{-1}$, which yields,
\begin{equation}
\frac{1}{\sqrt{r^2 + \beta^2}} =  \frac{1}{\beta\sqrt{1+ \frac{r^2}{\beta^2}}} = \frac{1}{\beta} (1 + \mathcal{O}(\tilde{x})),
\label{eq:SecondRatio}
\end{equation}						
where	$\tilde{x} = r/\beta < 1$. Proceeding as in Case 1, we obtain,
\begin{equation}
I = \frac{1}{\beta} (1 + \mathcal{O}(\tilde{w})),
\end{equation}
$|\tilde{w}| < 1$, $ \tilde{w} = \max\{ \tilde{x}, y\}$. This implies,  
\begin{equation}
\bar{V}_2 = -\frac{Gm_0m_2}{\beta}(1 + \mathcal{O}(\tilde{w})) = -\frac{Gm_0m_2}{\beta} + \mathcal{O}(m_0m_2).
\label{eq:PotCase2}
\end{equation}

\noindent Since $\beta$ is a constant, then in the derivation of $\hat{E}_{\tilde{n}}$, we replace $E$ with $E + (Gm_0m_2/\beta) + \mathcal{O}(m_0m_2)$ in (\ref{eq:RDE}) keeping $V_1$ as was used in the case of $m_2=0$ in (\ref{eq:RDE}).  This yields,
\begin{equation}
\hat{E}_{\tilde{n}} = - { {2\nu \pi^2 (Gm_0m_1)^2}\over{\sigma^2 \tilde{n}^2} } - \frac{Gm_0m_2}{\beta} +  \mathcal{O}(m_0m_2)
\label{eq:NewEnergyCase2}
\end{equation}
This can be reduced to,
\begin{equation}
\hat{E}_{\tilde{n}} = -\frac{4 \sigma}{\tilde{n}^2}   - \frac{Gm_0m_2}{\beta} + \mathcal{O}(m_0m_2).
\label{eq:NewEnergyCase2Simplified}
\end{equation}
\medskip\medskip\medskip

\noindent
Case 3: $r  = \beta$
\medskip

   In this final case,  
\begin{equation}
\frac{1}{\sqrt{r^2 + \beta^2}} =  \frac{1}{r\sqrt{1+ \frac{r^2}{\beta^2}}} = \frac{1}{\sqrt{2}r}.
\label{eq:ThirdRatio}
\end{equation}						
Thus, 
\begin{equation}
I =  \frac{1}{ \sqrt{r^2 + \beta^2}} \frac{1}{\sqrt{1 - \frac{2\beta r}{r^2 + \beta^2}\cos(\phi-\theta) }} = \frac{1}{\sqrt{2}r} \frac{1}{\sqrt{1 - h(\psi)}},
\label{eq:ratiocase3}
\end{equation}
where $h = \cos(\psi)$, $\psi = \phi-\theta$. We assume that $P_0$ does not collide with $P_2$, implying $ \psi \neq 0, \pm 2j \pi$, $j = 1,2,3,...$.  Thus, $|h| < 1$. We can write $I$ as,
\begin{equation}
I = \frac{1}{\sqrt{2}r} (1 + \mathcal{O}(h)).
\end{equation}                   
Hence, 
\begin{equation}
\bar{V}_2 = -\frac{Gm_0m_2}{\sqrt{2}r} + \mathcal{O}(m_0m_2).
\label{eq:PotCase3}
\end{equation}
Proceeding as in Case 1, 
\begin{equation}
\hat{E}_{\tilde{n}} = - {{ 2\nu \pi^2 (Gm_0(m_1 + \frac{1}{\sqrt{2}}m_2))^2}\over{\sigma^2 \tilde{n}^2} } + \mathcal{O}(m_0m_2)
\label{eq:NewEnergyCase3}
\end{equation}
This can be written as,
\begin{equation}
\hat{E}_{\tilde{n}} = -\frac{4 \sigma}{\tilde{n}^2}(1 + \frac{1}{\sqrt{2}}\mu + \frac{1}{2}\mu^2)+ \mathcal{O}(m_0m_2).
\label{eq:NewEnergyCase3Simplified}
\end{equation}    


\medskip\medskip\medskip\medskip




\section*{References}

\begin{thebibliography}{99}

\bibitem{SiegelMoser:1971} 
  C.~L.~Siegel and J.~K.~Moser,  
	{\em Lectures on Celestial Mechanics}, Springer Verlag, Grundlehren Series, Heidelberg-Berlin, 1971.
	
\bibitem{Lanczos:1949}
C. ~Lanczos, {\em The Variational Principals of Physics}, University of Toronto Press, 1949.

\bibitem{Pollard:1976}
H. ~Pollard,  {\em Celestial Mechanics}, The Carus Mathematical Monographs, Mathematical Association of America, no. 13, Washington, D.C., 1976.

\bibitem{Stiefel:1971}
E.L. ~Stiefel; G. Scheifele, {\em Linear and Regular Celestial Mechanics}, {\bf 174}, Springer-Verlag, New York, 1971.


\bibitem{Szebehely:1967}
V. ~Szebehely,{\em Theory of Orbits}, Academic Press, New York, 1967.

\bibitem{Moulton:1970}
F.R.~Moulton, {\em An Introduction to Celestial Mechanics}, Macmillan, London, 1917. Reprinted by Dover, New York, 1970.



%\bibitem{TaylorFrench:1979}
%A.P.~French, E.W.~Taylor, {\em An Introduction to Quantum Physics}, MIT Introductory Physics Series, CRC Press, 1979.

\bibitem{BelbrunoMarsden:1997}
E.~Belbruno, B.~Marsden,
% "Resonance hopping in comets", Astronomical J., 
{\em Astron. J.}, \ {\bf 113}, 1433-1444 (1997).

\bibitem{Ohtsuka:2008}
K.~Ohtsuka; T~Ito, M.~Yoshikawa; D.J.~Asher; H.~Arakida, 
% “Quasi-Hilda Comet 147P/Kushida-Muramatsu:Another long temporary satellite capture by Jupiter”,
{\em Astronomy and Astrophysics}, {\bf 489}, 1355-1362(2008).

\bibitem{Naidon:2017}
P.~Naidon; S.~Endo, 
%“Efimov physics: a review”, 
{\em Rep. Prog. Phys.},{\bf 80}, 056001(2017).



\bibitem{Sommerfeld:1921}
A.~Sommerfeld,
{\em Atombau und Spektrallinien}, Friedr. Vieweg \& Sohn, Braunschweig, 1921.

\bibitem{KLMR:2000}
W.S.~Koon, M.W.~Lo, J.E.~Marsden, S.D.~Ross, 
%"Heteroclinic connections between periodic orbits and resonance transitions in celestial mechanics", 
{\em Chaos}, \ {\bf 10}, 427-469 (2000).

\bibitem{KLMR:2001}
W.S.~Koon, M.W.~Lo, J.E.~Marsden, S.D.~Ross, 
%\rq\rq{}Resonance and Capture of Jupiter Comets", 
{\em Celest. Mech. Dyn. Astron.}, \ {\bf 81}, 27-38 (2001).



\bibitem{BTG:2008}
E.~Belbruno, F.~Topputo, M.~Gidea,
% "Resonance transitions associated with weak capture in the restricted three-body problem", 
{\em Advances \ in \ Space \ Research}, \ {\bf 42}, no. 8, 1330-1352 (2008).

\bibitem{Diosi:1984}
L.~Di\'osi, 
%"Gravitation and quantum-mechanical localization of macro-objects",
{\em Physics \ Letters \ A}, \ {\bf 105}, 199-202(1984).

\bibitem{Penrose:1996}
%"On gravity's role in quantum state reduction",
R.~Penrose, {\em Gen.\ Relativity \ and \ Gravitation}, \ {\bf 28} , Issue 5, 581-600(1996).

\bibitem{Belbruno:1997}
E.~Belbruno, 
%"Fast resonance shifting as a mechanism of dynamic instability illustrated by comets and CHE trajectories", 
{\em Annals \ New \ York \ Academy \ of \ Sciences},  in {\em Near Earth Objects}(ed. J. Remo), \ {\bf 822} 195-225 (May 1997).




\bibitem{Belbruno:2004}
  E.~Belbruno,
  {\it Capture Dynamics and Chaotic Motions in Celestial Mechanics},
  Princeton University Press, 2004.


\bibitem{Kummer:1979} 
	M.~Kummer,
%"On the stability of Hill's solutions in the plane restricted three body problem", 
{\it Am.\ J.\ Math. } \ {\bf 101}, 1333-1354 (1979).

\bibitem{BGT:2010}
E.~Belbruno, M.~Gidea, F.~Topputo, 
%"Weak stability boundary and manifolds", 
{\em SIAM \ J.\ Appl.\ Dyn.\ Sys.}, \ {\bf 9}, 1061-1089 (2010).

\bibitem{BGT:2013}
E.~Belbruno, M.~Gidea, F.~Topputo, 
%"Geometry of the Weak Stability Boundaries", 
{\em Qual. \ Theory\ Dyn..\ Sys.}, \ {\bf 12}, 53-66 (2013).


\bibitem{GarciaGomez:2007}
% "A notedon weak stability boundaries"
G.~Garcia, G.~Gomez, {\em Cel. \ Mech. \ Dyn. \ Astr.}, \ {\bf 97}, 87-100(2007). 

\bibitem{TB:2009}
F.~Topputo, E.~Belbruno,
%"Computation of weak stability boundaries: Sun-Jupiter",
{\em Cel. \ Mech. \ Dyn. \ Astr.}, \ {\bf 105}, 3-17(2009) 


\bibitem{Belbruno:1987}
E.~Belbruno,
%"Lunar capture orbits, a method of computing earth-moon trajectories and the Lunar GAS mission", 
{\em Proceedings \ of \ the \ AIAA/DGLR/JSASS \ Inter. \  Elec. \  Propl. \ Conf.}, no. 87-1054 (1987).

%\bibitem{Belbruno:2007}
%E.~Belbruno,
%{\it Fly Me to the Moon},
%Princeton University Press, 2007.	

\bibitem{BelbrunoMiller:1993}
E.~Belbruno, \ J.~Miller
%"Sun-perturbed earth-to-moon transfers with ballistic capture",
{\em J. Guid.\ Control \ Dyn. \ Astr.}, \ {\bf 16}, 770-775 (1993).

%\bibitem{BG:2005}
%"Where did the moon come from?",
%E.~Belbruno, J.R.~Gott III, {\em Astron. J.}, \ {\bf 129}, 1724-1745 (2005). 

\bibitem{BMMS:2012}
E.~Belbruno, A.~Moro-Martin, R.~Malhotra, D.~Savransky, 
%"Chaotic exchange of solid material between planetary systems: Implications for lithopanspermia"
{\em Astrobiology}, {\bf 12}, 1-21 (2012).

\bibitem{LMS:1985}
J.~Llibre, R.~Martinez, C.~Simo,
%"Transversality of the invariant manifolds associated to the Lyapunov family of periodic orbits near L2 in the restricted three-body problem",
{\em J. \ Diff. \ Equ.}, \ {\bf 58}, 104-156(1985).

% EARLY AIAA PAPER


%OR MY EARLIER PAPER

\bibitem{Conley:1968}
C.~Conley, 
%"Low energy transit orbits in the restricted three-body problem",
{\it SIAM \ J.\ Appl.\ Math.}, \ {\bf 16}, 732-746 (1968).


%\bibitem{Sakurai:1984}
%J.J. ~Sakurai,  {\em Modern Quantum Mechanics}, Addison-Wesley, 1994. 
                
\bibitem{Atkins:2005}
P.~Atkins, R.~Friedman, {\em Molecular Quantum Mechanics}, Fourth Edition, Oxford University Press, 2005.
            
\bibitem{Abramowitz:1974}
M.~Abramowitz, I.~Stegun,  {\em Handbook of Mathematical Functions}, Dover, 1974. 





\end{thebibliography}
\end{document}